\documentclass[12pt,titlepage]{article}
\usepackage{epsfig}
\usepackage{graphicx}
\usepackage{amsmath}

\def\beq{\begin{equation}}
\def\eeq{\end{equation}}
\def\bea{\begin{eqnarray}}
\def\eea{\end{eqnarray}}
\def\bean{\begin{eqnarray*}}
\def\eean{\end{eqnarray*}}
\def\dCSA{{N}}
\def\J{{j}}
\def\Inertia{{\cal I}}
\def\X{{t}}
\def\AllOne{\mathbf Q}

\begin{document}

\title{On the thermodynamic stability of rotating black holes in higher dimensions --- 
a comparison of thermodynamic ensembles}

\author{{Brian P. Dolan}\\
{\em Department of Mathematics, Heriot-Watt University}\\ 
{\em Colin Maclaurin Building, Riccarton, Edinburgh, EH14 4AS, U.K.}\\
[5mm]
{\em Maxwell Institute for Mathematical Sciences, Edinburgh, U.K.}\\
[5mm]
{\em Email:} {\tt B.P.Dolan@hw.ac.uk}
}

\maketitle
\begin{abstract}

\bigskip

Thermodynamic potentials relevant to the microcanonical, the canonical and the 
grand canonical ensembles, associated with rotating black holes 
in $D$-dimensions, are analysed and compared. 

Such black holes are known to be thermodynamically unstable, 
but the instability is a consequence of a subtle interplay between specific heats and the moments
of inertia and it manifests itself differently in the different ensembles. 
A simple relation between the product of the 
specific heat and the determinant of the 
moment of inertia in both the canonical and the grand canonical ensembles is derived.

Myers-Perry black holes in arbitrary dimension are studied in detail.
All temperature extrema in the microcanonical ensemble are determined and
classified.  The specific heat and the moment of inertia tensor are evaluated
in both the canonical and the grand canonical ensembles in any dimension. 
All zeros and poles of the specific heats, as a function of the angular momenta, 
are determined and the eigenvalues of the isentropic moment of inertia tensor are studied and classified.

It is further shown that many of the thermodynamic properties of a Myers-Perry black hole in $D-2$ dimensions can be obtained from those of a black hole in $D$ dimensions by sending one of the angular momenta to infinity.

\bigskip
\noindent

\begin{flushright}
\end{flushright}

\end{abstract}

\section{Introduction}

An immediate consequence of Hawking's famous result that Schwarzschild black holes
in four dimensions have a temperature that is inversely proportional to their mass 
\cite{Hawking}
is that any such black hole is thermodynamically unstable,
due to a negative specific heat.  They can however be stabilised by putting them in
a box with thermal walls \cite{York} or by introducing a negative cosmological
constant of sufficient magnitude \cite{HawkingPage}.  While the specific heat of a Schwarzschild black hole can be
rendered positive by making it rotate sufficiently fast
this does not stabilise the black hole as the moment of inertia tensor then
becomes negative, maintaining the instability.

The generalisation of the Kerr metric to a class of rotating black holes in $D$-dimensions, found by Myers and Perry \cite{MP}, provides an arena for testing these ideas in a more general context. 
In 4-dimensions there is a maximum angular momentum that a rotating 
black hole can sustain, corresponding to an extremal black hole with vanishing
Hawking temperature, but in higher dimensions this is not the case.
There is more than one angular momentum in $D>4$, corresponding to the fact that the rank of $SO(D-1)$ is greater than one for $D>4$, and some, but not all,
of the angular momenta can become arbitrarily large --- the phenomenon of ultra-spinning black holes \cite{MP}.  Infinite momentum does not however imply infinite angular velocity, rather the corresponding angular velocity vanishes as an angular momentum diverges --- the infinite angular momentum is due to a singularity in the moment of inertia of the black hole and is not due to infinite angular velocity.

It was suggested some time ago that
there should be  a link between the thermodynamic properties of black holes,
in particular the second law of thermodynamics, 
and dynamical instability for $D>4$, \cite{GL}.
Stability of Myers-Perry black holes
was analysed in \cite{EM} and an extensive literature on the
subject of the thermodynamic and dynamical instability of 
rotating black holes in higher dimensions
has since emerged \cite{KLR}-\cite{DMS}.
In particular it has been shown, with a very general argument 
utilising only the Smarr relation and the first law,  
that all asymptotically flat electrically neutral 
solutions of the vacuum Einstein equations in $D$-dimensions
are thermodynamically unstable, \cite{DFMR}.
 Nevertheless it is still instructive to examine the details of thermodynamic stability in the different ensembles and in specific cases.

The thermodynamic quantities of interest are the mass (internal energy),\footnote{When there is a non-zero cosmological constant it is argued in \cite{KRT} that the
mass is more correctly thought of as the enthalpy rather than the internal
energy.  These are of course the same for zero pressure and in this work
we  make no distinction between enthalpy and internal energy.}
the entropy, the temperature, the angular momenta and the angular velocities.
With Newton's constant $G_N=1$ and $c=1$ these can all be given dimensions of length to
some power:

\medskip

\setlength{\tabcolsep}{15pt}
\centerline{\begin{tabular}{| l| c| c| }
\hline
 Mass, $M$   &  $D-3$ \\
  Entropy, $S$ (area)  & $D-2$ \\
  Angular momenta, $J^i$ & $D-2$ \\
 Temperature,  $T$ & $-1$ \\
 Angular velocity,  $\Omega_i$ & $1$ \\
\hline
\end{tabular}}

\medskip 

\noindent It is thus natural to consider $M$, $S$, and $J^i$ to be extensive while
$T$ and $\Omega_i$ are intensive, and this classification will be adopted here.

In \S\ref{sec:EnsembleRelation} a general relation between the canonical and the grand canonical ensembles, for electrically neutral rotating
black holes, is derived.
For the canonical ensemble the Hessian, $\partial ^2 U$, of the internal
energy $U(J^i,S)$ is shown to have determinant
\beq \det(\partial^2 U)=\frac {1}{\beta C_J \det\Inertia_T}\eeq
where $C_J$ is the specific heat at constant angular momentum and $\Inertia_T$
is the isothermal moment of inertia tensor.
For the grand canonical ensemble, on the other hand, the Hessian 
 $\partial ^2 G$ of the grand canonical potential, $G(\Omega_i,T)$, is shown to satisfy
\beq \det(-\partial^2 G)= \beta C_\Omega \det\Inertia_S,
\label{detG}
\eeq
where $C_\Omega$ is the specific heat at constant angular velocity
and $\Inertia_S$  is the isentropic moment of inertia tensor.
Standard thermodynamics arguments then imply that
\beq
C_J \det\Inertia_T = C_\Omega \det\Inertia_S,
\label{eq:detITdetIS}
\eeq 
which is one of  our main results.

We then compare the thermodynamics of  Myers-Perry black
holes in the different ensembles.
In one of the first papers on the stability of Myers-Perry black holes 
\cite{EM} it was observed that, as one of the angular momenta is increased keeping the others zero and the mass fixed (the microcanonical ensemble), 
there is a minimum in the temperature.  The authors suggested that this was a signal of an instability --- that there should be dynamical negative modes leading to a more stable solution of Einstein's equations, but with less symmetry.  
This gave further support, beyond the non-rotating case studied in \cite{GL},
 to the idea that thermodynamic and dynamical instability of
black holes in higher dimensions are intimately related.
Hints of the instability can be seen in the microcanonical ensemble
in which the entropy is a monotonically
decreasing concave function of angular momenta, at constant mass, 
until the temperature hits a minimum
at which point the entropy has an inflection point and becomes concave in at least one direction \cite{EHNOR}. 

Thermodynamic instability manifests itself in different ways in the various ensembles.   In the grand canonical ensemble the
grand canonical potential $G(\Omega_i,T)$ is considered as a function of
intensive variables and thermodynamic stability requires that $G$ be a totally concave function of its arguments \cite{Callen}. The particular cases of asymptotically flat Kerr and Myers-Perry
black holes were investigated in \cite{MPS} for $4\le D \le 6$ and it was
shown that the specific heat $C_J$
is negative when all angular momenta vanish, but can become positive when some
of the angular momenta become large enough.   However when the angular momenta are large enough for the specific heat to be positive
 the isothermal moment of inertia, all of whose eigenvalues are positive for zero rotation,
has at least one negative eigenvalue --- there is thus always an instability. 
One of the results of the present work is to
extend the explicit analysis of \cite{MPS} to all $D$ and show that the same phenomenon persists. 

The relationship between the microcanonical and the grand canonical
ensembles was studied in \cite{ART} and
in \S\ref{sec:anyD} we extend this analysis further and
derive a number of relations between 
the temperature, the specific heat at constant angular velocity
$C_\Omega$, and the isentropic moment of inertia tensor $\Inertia_S$
for Myers-Perry black holes in any dimension.
We show that there is a branched hypersurface in angular momentum space where 
$\beta C_\Omega$ (with $\beta = \frac 1 T$) develops a pole and that 
this is the same 
hypersurface as the one on which the temperature is minimised in the microcanonical ensemble.
This hypersurface can be obtained from the extremal $T=0$ 
hypersurface by analytically continuing $(J^i)^2$ to $-(J^i)^2$,
keeping the entropy constant.
There is yet another significant hypersurface, one with a number of branches on which $\beta C_\Omega$ vanishes,
and on this hypersurface the isentropic moment of inertia tensor develops an infinite 
eigenvalue, in the form of a pole. This pole exactly cancels the zero
in $\beta C_\Omega$  in the Hessian
of the grand canonical potential. The branches of this hypersurface divide the space
of angular momenta into separate regions determined by the signature of $\Inertia_S$.

A by product of our analysis is that the thermodynamic properties of a Myers-Perry
black hole in $D-2m$ dimensions, in the micro or the grand canonical ensemble,
can be obtained from those of a Myers-Perry black hole in
$D$ dimensions by sending $m$ of the angular momenta to infinity in the latter.

In \S\ref{sec:thermodynamics} the thermodynamics of rotating black holes 
in the different ensembles are analysed and equation (\ref{eq:detITdetIS}) derived along with other relations between the various thermodynamic quantities.
In \S\ref{sec:MP} Myers-Perry black holes are studied 
and it is shown explicitly how
the specific heats and moments of inertia conspire to satisfy the general
relations of \S\ref{sec:thermodynamics}. The results are summarised in \S\ref{sec:conclusions} and some technical results required
in the analysis are relegated to five appendices.

\section{Thermodynamics of rotating black holes\label{sec:thermodynamics}}

Rotating black holes in $D>4$ space-time dimensions must be treated slightly
differently for even and odd $D$
because the rotation group  $SO(D-1)$
has different characterisations 
of angular momenta in the even and odd 
dimensional cases.
The Cartan sub-algebra has dimension $\frac{D-2}2$ for even $D$ and $\frac{D-1}2$ for odd $D$ so a general state of rotation is specified by 
$\frac{D-2}2$ independent angular momenta in even $D$ and $\frac{D-1}2$
in odd $D$.  Let $\dCSA=\left\lfloor \frac{D-1}2 \right\rfloor$, the integral part
of $\frac{D-1}2$, be the dimension of the Cartan sub-algebra of $SO(D-1)$,
then there are $\dCSA$ independent angular momenta $J^i$,
$i=1,\ldots,\dCSA$.  It is notationally convenient to introduce a parameter 
$\epsilon=\bigl(1+(-1)^D\bigr)/2$ in terms of which 
\beq \dCSA=\frac{D-1-\epsilon}2.\eeq

In the microcanonical ensemble the energy is fixed and we
chose as thermodynamic control parameters the extensive quantities,
$J^i$ and $M$, with the
entropy $S(J^i,M)$ being the thermodynamic potential, which is convenient
for differentiation keeping $M$ fixed.
In the canonical ensemble the energy is allowed to fluctuate 
and the internal energy 
\beq U(J,S)=M\eeq
is used as the thermodynamic potential.
In the grand canonical ensemble all extensive variables are allowed to fluctuate and the intensive variables are used as control parameters, 
the relevant thermodynamic potential is then the grand canonical potential 
\beq G(\Omega,T)=U - TS -  \Omega_i J^i.\label{eq:Grand}\eeq

The grand canonical potential is related to the Euclidean formulation, since the
Euclidean action $I_E$ is related to the mass by a Legendre transform
\beq
T I_E = M - TS -\Omega_i J^i \qquad \Rightarrow \qquad T I_E=G(\Omega,T).
\eeq

\subsection{Microcanonical ensemble}

For completeness we summarise in this sub-section
some of the results pertaining to the microcanonical analysis in \cite{DFMS}.

With the entropy expressed as a function of $J^i$ and $M$,  $S(J,M)$,
we have
\beq 
\left.\frac{\partial S}{\partial J^i}\right|_M = -\beta \Omega_i,
\qquad
\left.\frac{\partial S}{\partial M}\right|_{J^i} = \beta,
\eeq
where $\beta=\frac 1 T$.  These thermodynamic quantities have a geometrical
interpretation in the Euclidean formulation of the black hole, where demanding
absence of a conical singularity requires periodicity in imaginary time,
$\tau=it$, with $\tau$ identified with $\tau + \beta$, and 
periodicity in imaginary angle \cite{DFMS}, $\varphi_i = i\phi_i$, with 
$\varphi_i$ identified with $\varphi - \beta\Omega_i$. Thus the 1-form $dS$
determines the size of the $(\tau,\varphi_i)$ torus,
\beq \tau dM + \varphi_i d J^i \sim \tau dM + \varphi_i d J^i + d S.
\eeq
 
For
thermodynamic stability the entropy should be purely concave \cite{Callen}, 
which requires
that the Hessian $H_{AB}=-\frac{\partial ^2 S}{\partial \tilde x^A \partial \tilde x^B}$, where $\tilde x^A=(J^i,M)$, must be a positive definite matrix. 
The identity
\beq 
\det \bigl(H_{AB}\bigr)= -\frac{1}{(D-3)MT} \det\bigl(H_{ij}\bigr),
\eeq
for asymptotically flat black holes,
is derived in \cite{DFMR}, where $H_{ij}$ is the $\dCSA\times\dCSA$ matrix
\beq \label{eq:Hijdef}
H_{ij} =-\left(\frac{\partial ^2 S}{\partial J^i \partial J^j}\right)_M.
\eeq
One can immediately conclude that such black holes can never be thermodynamically stable in $D\ge 4$, since $\det (H_{AB})>0$ requires $\det(H_{ij})<0$, hence at least
one eigenvalue of $H_{ij}$ would have to be negative and $S$  cannot be a concave function.

\subsection{Canonical ensemble \label{sec:canonical}}

The canonical ensemble uses the internal energy $U(J^i,S)$ as
thermodynamic potential, depending on extensive arguments that are the Legendre
transforms of $\Omega_i$ and $T$, and for black holes $U$ is identified with
the ADM mass, at least in the asymptotically flat case.  
Stability requires that $U$ be a totally convex function of its arguments.
Let $x^A=(J^i,S)$, with $A=1,\ldots,N+1$, $x^i=J^i$ and $x^{N+1}=S$,
then 
\beq
T=\left.\frac{\partial U}{\partial S}\right|_J,\qquad
\Omega_i=\left.\frac{\partial U}{\partial J^i}\right|_S
\eeq
and
$U_{AB}=\frac{\partial^2 U}{\partial x^A \partial x^B}$ must be a positive matrix.
Explicitly
\beq 
 U_{AB}=\begin{pmatrix}
\frac{\partial^2 U}{\partial J^i \partial J^j}
&  \frac{\partial^2 U}{\partial J^i \partial S}\\
 \frac{\partial^2 U}{\partial S \partial J^j } &  
\frac{\partial^2 U}{\partial S^2}
\end{pmatrix}
=
\begin{pmatrix}
\bigl({\Inertia_S}^{-1}\bigr)_{ij}
&\zeta_i  \\
 \zeta_j &  
(\beta C_J)^{-1}
\end{pmatrix},\label{eq:UHessian}
\eeq
where the symmetric matrix
 \beq
{\Inertia_S}^{ij}=\left.\frac{\partial J^i}{\partial \Omega_j}\right|_S
\eeq
is the isentropic moment of inertia tensor;
$C_J$ is the specific heat at constant $J$,
\beq C_J=\left(\frac{\partial U}{\partial T}\right)_J
=T\left(\frac{\partial S}{\partial T}\right)_J\,
\eeq
and
\beq \zeta_i= \left.\frac{\partial \Omega_i}{\partial S}\right|_J
=\left.\frac{\partial T}{\partial J^i}\right|_S
\eeq
(this last equation is a Maxwell relation).

A necessary, but not sufficient, condition for stability is thus
\beq \det(\partial^2 U) = \frac
{\det\bigl({\Inertia_S}^{-1} - \beta C_J \zeta \zeta^T\bigr)}{\beta C_J} >0. \label{eq:Ustability}
\eeq

\subsection{Grand canonical ensemble}

In the grand canonical ensemble stability requires that $G(\Omega,T)$ be a concave function.
Let $y_A=(\Omega_i, T)$, with $y_i=\Omega_i$ and $y_{N+1}=T$,
then 
\beq
S=-\left.\frac{\partial G}{\partial T}\right|_\Omega,\qquad
J^i=-\left.\frac{\partial G}{\partial \Omega_i}\right|_T
\eeq
and
$G^{AB}=\frac{\partial^2 G}{\partial y_A \partial y_B}$ must be a negative matrix.
Explicitly
\beq 
 -G^{AB}=\begin{pmatrix}
-\frac{\partial^2 G}{\partial \Omega_i \partial \Omega_j}
&  -\frac{\partial^2 G}{\partial \Omega_i \partial T}\\
 -\frac{\partial^2 G}{\partial T \partial \Omega_j } &  
-\frac{\partial^2 G}{\partial T^2}
\end{pmatrix}
=
\begin{pmatrix}
{\Inertia_T}^{ij}
&\eta^i  \\
 \eta^j &  
\beta C_\Omega
\end{pmatrix}, \label{eq:GrandHessian}
\eeq
where the symmetric matrix
 \beq
\Inertia_T^{ij}=\left.\frac{\partial J^i}{\partial \Omega_j}\right|_T
\eeq
is the isothermal  moment of inertia tensor;
$C_\Omega$ is the specific heat at constant $\Omega$,
\beq C_\Omega
=T\left(\frac{\partial S}{\partial T}\right)_\Omega
=
-T\left(\frac{\partial^2 G}{\partial T^2}\right)_\Omega\,
\eeq
and
\beq \eta^i= \left.\frac{\partial J^i}{\partial T}\right|_\Omega
=\left.\frac{\partial S}{\partial \Omega_i}\right|_T.\label{eq:etaMaxwell}
\eeq

A necessary, but not sufficient, condition for stability is thus
\beq \det(-\partial^2 G) = \beta C_\Omega 
\det\bigl(\Inertia_T - (\beta C_\Omega)^{-1} \eta \eta^T\bigr)>0. 
\label{eq:Gstability}
\eeq

\subsection{Relation between the canonical and the grand canonical
ensembles \label{sec:EnsembleRelation}}

The canonical and the grand canonical ensembles are of course related.
An immediate consequence of the Legendre transform (\ref{eq:Grand}) is that
\beq -G^{AB} = (U^{-1})^{AB} \label{eq:LegendreInverse}\eeq
and this has important consequences for the individual components. 

A relation between the specific heats was derived in \cite{MPS},
\bea
\left.\frac{\partial S}{\partial T}\right|_\Omega &=& 
\left.\frac{\partial S}{\partial T}\right|_J
+ \left.\frac{\partial S}{\partial J^j}\right|_T
\left.\frac{\partial J^j}{\partial T}\right|_\Omega
=
\left.\frac{\partial S}{\partial T}\right|_J
+ \left.\frac{\partial S}{\partial \Omega_i}\right|_T
\left.\frac{\partial \Omega_i}{\partial J^j}\right|_T
\left.\frac{\partial J^j}{\partial T}\right|_\Omega\\
&=&
\left.\frac{\partial S}{\partial T}\right|_J
+ \bigl(\Inertia_T^{-1}\bigr)_{ij}\eta^i \eta^j\\
&\Rightarrow& \qquad \beta C_\Omega = \beta C_J + \bigl(\Inertia_T^{-1}\bigr)_{ij}\eta^i \eta^j,\label{eq:CJOmegaRelation}
\eea
where the Maxwell relation (\ref{eq:etaMaxwell}) has been used.

Similar manipulations can be used to relate the isentropic and isothermal
moment of inertia tensors,
\beq
{\Inertia_T}^{ij} = {\Inertia_S}^{ij} +(\beta C_\Omega)^{-1} \eta^i\eta^j.  
\eeq
Or equivalently,
\beq
\bigl({\Inertia_S^{-1}}\bigr)_{ij} = \bigl({\Inertia_T^{-1}}\bigr)_{ij} 
+\beta C_J \zeta_i\zeta_j.  
\eeq

The stability conditions (\ref{eq:Ustability}) and (\ref{eq:Gstability}) 
can thus be expressed as
\beq
\det(\partial^2 U)=\frac {1} {\beta C_J \det(\Inertia_T)}>0
\label{eq:d2U}\eeq
and 
\beq
\det(-\partial^2 G)=\beta C_\Omega \det(\Inertia_S)>0.
\label{eq:minusd2G}
\eeq

Equation (\ref{eq:LegendreInverse}) now gives the identity
\beq 
\beta C_J \det(\Inertia_T)=\beta C_\Omega \det(\Inertia_S).
\label{eq:equaldets}\eeq
A new instability would be expected to develop every time one 
of the eigenvalues of $-\partial^A \partial^B G$ 
changes from a positive to a negative
value, either by going through zero or infinity. In general 
this might be expected to happen on a hypersurface on which 
$\det(-\partial^2 G)$ is either zero or infinity, but we shall see that, at least in the case of Myers-Perry black holes, there are some subtle cancellations
so that the change is not reflected in the determinant.

The form of the Hessians (\ref{eq:UHessian}) and (\ref{eq:GrandHessian}) 
can be simplified by using a Legendre transform on the scalar
variable, respectively $S$ and $T$.
In the canonical ensemble let $x^{A'}=(J^i,T)$ and
\beq F(J^i,T)= U - TS. \eeq
Then the co-ordinate transformation matrix is 
\beq
\frac{\partial x^{A'}}{\partial x^B}= 
\begin{pmatrix}
\delta^i{}_j & 0   \\
 \zeta_j &  
(\beta C_J)^{-1}
\end{pmatrix}
\eeq
with inverse
\beq
\frac{\partial x^A}{\partial x^{B'}}= 
\begin{pmatrix}
\delta^i{}_j & 0   \\ -\beta C_J\zeta_j &  
\beta C_J
\end{pmatrix}.
\eeq
So, in $(J^i,T)$ co-ordinates,
\beq
 U_{A'B' }=U_{CD}\frac{\partial x^C}{\partial x^{A'}}
\frac{\partial x^D}{\partial x^{B'}}
=
%\begin{pmatrix}
%\bigl( \Inertia_S^{-1}\bigr)_{ij} - (\beta C_J) \zeta_i\zeta_j &  0 \\
%0&  \beta C_J
%\end{pmatrix}=
\begin{pmatrix}
\bigl( \Inertia_T^{-1}\bigr)_{ij}  &  0 \\
0&  \beta C_J
\end{pmatrix}.\label{eq:diagd2S}
\eeq
is partially diagonalised.\footnote{Of course
$U_{A'B' } \ne {\partial^2 F}/{\partial x^{A'}\partial x^{B'}}$.
The stability properties of the canonical ensemble are determined by the
signature of the Hessian  $U_{AB}$, which is the same as 
that of $U_{A'B'}$. The matrix ${\partial^2 F}/{\partial x^{A'}\partial x^{B'}}$
has a different signature.}

Similarly, in the grand canonical ensemble, we can transform from
$y_A=(\Omega_i,T)$ to $y_{A'}=(\Omega_i,S)$ to get
\beq
 -G^{A'B' }=-G^{CD}\frac{\partial x^{A'}}{\partial x^C}
\frac{\partial x^{B'}}{\partial x^D}  =
%\begin{pmatrix}
%\Inertia_T^{ij} - (\beta C_\Omega)^{-1} \eta_i\eta_j &  0 \\
%0&  (\beta C_\Omega)^{-1}
%\end{pmatrix}=
\begin{pmatrix}
\Inertia_S^{ij}  &  0 \\
0&  (\beta C_\Omega)^{-1}
\end{pmatrix}.\label{eq:diagd2G}
\eeq
Note that, although the canonical ensemble implicitly involves $\Inertia_S^{-1}$, its stability properties are most easily seen using $\Inertia_T^{-1}$
in (\ref{eq:diagd2S}) and, while the grand canonical ensemble implicitly involves $\Inertia_T$, its stability properties are most easily studied using $\Inertia_S$ in (\ref{eq:diagd2G}).

\section{Myers-Perry black holes \label{sec:MP}}

Myers-Perry black holes in $D$ space-time dimensions have an event horizon which
has the topology of a $(D-2)$-dimensional sphere.
This can be described in
terms of Cartesian co-ordinates ${\tt x}_a$ in
${\bf R}^{D-1}$ by
\beq \sum_{a=1}^{D-1} {\tt x}_a^2 =1,\eeq
and we can write this as
\beq \sum_{i=1}^\dCSA \rho_i^2 + \epsilon y^2=1,\eeq
where ${\tt x}_{2i-1}+i{\tt x}_{2i} = \rho_i e^{i\phi_i}$, $i=1,\ldots,\dCSA$, are complex co-ordinates
for both the even and odd cases while $y={\tt x}_{D-1}$ is only present for even $D$.

Then $\rho_i$, $\phi_i$ and $y$ are co-ordinates that can be used
to parameterise the sphere and, for the black hole, $J^i$ are angular
momenta in the $({\tt x}_{2i-1},{\tt x}_{2i})$-plane.\footnote{Hereinafter we shall not
distinguish between upper and lower indices, $J_i=J^i$.}

The Myers-Perry line element can be expressed as
\bea
d s^2&=&- dt^2 +\frac {2\mu }U
\left( dt-\sum_{i=1}^\dCSA  a_i \rho_i^2d\phi_i\right)^2
\nonumber\\
&& +\left(\frac {U }{Z-2\mu}\right)d r^2 + \epsilon\, r^2 d y^2 
 +\sum_{i=1}^\dCSA(r^2+a_i^2)(d\rho_i^2+\rho_i^2d\phi_i^2 ),
\nonumber\eea
where the functions $Z$ and $U$ are 
\bea 
Z&=&\frac{1}{r^{2-\epsilon}}\prod_{i=1}^\dCSA(r^2+a_i^2)\\
U&=& Z\left(1-\sum_{i=1}^\dCSA\frac{a_i^2\rho_i^2}{r^2+a_i^2} \right).
\nonumber
\eea
The $a_i$ are rotation parameters in the $({\tt x}_{2i-1},{\tt x}_{2i})$-plane and $\mu$
is a mass parameter.  We use units in which the $D$-dimensional Newton's constant and the speed of light are set to one.

There is an event horizon at $r_h$, the largest root of $Z-2\mu=0$, 
so
\beq 
\mu=\frac{1}{2r_h^{2-\epsilon}}\prod_{i=1}^\dCSA(r_h^2+a_i^2),
\label{eq:mudef}\eeq
and the area of the event horizon is
\beq 
{\cal A}_h=\frac{\varpi}{r_h^{1-\epsilon}}\prod_{i=1}^\dCSA(r_h^2+a_i^2)\,,
\eeq
Where $\varpi$ is 
is the volume of the round unit $(D-2)$-sphere,
\beq
\varpi= \frac{2\pi^{\frac{(D-1)}{2}}}{\Gamma\left(\frac{D-1}{2} \right)}\,.
\eeq

The Bekenstein-Hawking entropy is
\beq 
S=\frac{\varpi}{4 r_h^{1-\epsilon}}\prod_{i=1}^\dCSA(r_h^2+a_i^2)
\label{eq:Sdef}\eeq
and the Hawking temperature is
\beq
T=\frac{1}{4\pi r_h}\left(D-3-2\sum_{i=1}^\dCSA\frac{a_i^2}{r_h^2+a_i^2}\right).
%=\frac{r_h}{2\pi}\sum_{i=1}^\dCSA\frac{1}{r_h^2 + a_i^2} - \frac{(2-\epsilon)}{4\pi r_h}.
\label{eq:Tdef}
\eeq

The  angular momenta, the entropy and the ADM mass, $M$, of the black hole are related to each other, and to the metric parameters, via
\beq  J^i = \frac{2 M a_i}{D-2} = \frac{\mu\,\varpi a_i}{4\pi}, \qquad
M=\frac{(D-2)\varpi\mu}{8\pi}= \frac{(D - 2)S}{4 \pi r_h}\,,\label{eq:MJdef}\eeq
while the angular velocities are
\beq \Omega_i=\frac{a_i}{(r_h^2+a_i^2)}.\label{eq:Omegadef}\eeq

\subsection{Microcanonical ensemble}

The microcanonical ensemble was developed for Myers-Perry black holes in 
\cite{DFMSE}, \cite{DFMR} and \cite{DFMS}. 
In particular the Hessian (\ref{eq:Hijdef}) was evaluated explicitly
in \cite{DFMS} and is reproduced in appendix \ref{app:T_ext},
\beq 
H_{ij}=\frac{(D-2)} {2 r_h^2 T M} \left\{
\frac{(1-\J_i^2)}{(1+\J_i^2)^2}\delta_{ij}\right.
 \left.+ \frac{\omega_i \omega _j}{\pi r_h T}
\left(\frac {1}{1+\J_i^2}  + \frac {1}{1+\J_j^2}  -\frac 1 2 
+ \frac{2}{\X}\Omega^2 \right)
\right\},
\label{eq:HijMP}
\eeq
where $\J_i=\frac{2\pi J_i}{S}$, $\omega_i=\frac{\J_i}{1+\J_i^2}$
are dimensionless angular velocities, and $\Omega^2=\sum_{i=1}^\dCSA \omega_i^2$.

Black hole thermodynamics in $D>4$  has a subtle relation with dynamical
instability.
It was noted in \cite{EM} that, for $D\ge 6$, the temperature of a
Myers-Perry black hole, with only one $J^i\ne 0$, has a minimum as the
spin increases at fixed mass. 
Taking $J^1\ne 0$ and $J^i=0$ for $i=2,\ldots,\dCSA $
the minimum is at $\frac{a_1^2}{r_h^2}=\frac{D-3}{D-5}$ and in \cite{EM} it
was suggested that this minimum 
signals the onset of a dynamical instability for a rotating black hole. 
Thermodynamic functions are thus giving hints of 
possible dynamical instability and this was studied in  
\cite{DFMR} and \cite{DFMS}, where some
special cases of non-zero spin were analysed. These authors studied the
matrix (\ref{eq:HijMP}) in the symmetric cases were 
the non-zero $a_i$ are all equal,
\beq
a_1 =  \ldots = a_n =a \ne 0,\qquad   a_i=0 \quad \hbox{for}\quad  i=n,\ldots,\dCSA.
\label{eq:eq_a}\eeq
The entropy and the temperature 
decrease as the angular momenta are increased, at constant $M$, 
until the temperature reaches a minimum, and at precisely that
point the matrix $H_{ij}$ develops a zero eigenvalue, signalling the fact 
that the entropy ceases to be concave in that direction. 
The temperature has a minimum for configurations of the form (\ref{eq:eq_a}) with 
\beq \frac{a^2}{r_h^2}= \frac{D-3}{D-3-2n}.\label{eq:Tmin}\eeq
In particular $n=1$ gives the original expression in \cite{EM},
and the two special cases $n=1$ and $n=\dCSA$ were considered in \cite{DFMS},
while (\ref{eq:Tmin}) for general $n$ appeared in \cite{ART}. 

In appendix \ref{app:T_ext} all extrema of the temperature, 
in the microcanonical ensemble with fixed mass, 
\beq \left.\frac{\partial T}{\partial J^i}\right|_{M,\,{\mathbf J}_*}=0, \eeq
are found and classified.
For finite $J^i$ they are all of the same from as
(\ref{eq:eq_a}),
\beq J^1=\cdots = J^n=J_*, \qquad J^{n+1}=\cdots =J^\dCSA=0, \eeq
(up to permutations of the $J^i$) 
with 
\beq
\J_*^2 = \frac{D-3}{D-3-2n,}
\label{eq:Jstar}
\eeq
where $\J_*=\frac{2\pi J_*}{S}$ is the dimensionless angular momentum, in units 
of entropy.

The value of the temperature at the extrema (\ref{eq:Jstar})
is
\beq
T_*=\frac{1}{4\pi r_h}\frac{(D-3)(D-3-2n)}{(D-3-n)}.
\label{eq:extremumT}
\eeq
The temperature is a maximum, $T_{max}=\frac{D-3}{4\pi r_h}$, 
for non-rotating Schwarzschild-Tangherlini black holes ($n=0$).
For finite $J^i$ the stationary points ${\bf J}_*$ are saddle points 
of the temperature with minima along the directions
${\mathbf J}_*=(J_*,\cdots,J_*,0,\ldots,0)$ satisfying (\ref{eq:Jstar})
and maxima in the directions orthogonal to these. 
At the same time $H_{ij}$ in equation (\ref{eq:HijMP})
develops a single zero eigenvalue at ${\mathbf J}_*$,
in the direction ${\mathbf J}_*$, indicating an inflection point in
that direction.\footnote{It is argued in \cite{DFMSE} that this point of inflection is not in itself necessarily
a sign of dynamical instability: it can indicate a zero mode, taking
one Myers-Perry solution into another, 
rather than a negative mode dynamical instability.}
There are also $(n-1)$ negative eigenvalues of $H_{ij}$ at ${\mathbf J}_*$, 
(\ref{eq:Hijeigenvalues}), indicating convexity of the entropy in these directions with associated thermodynamic instabilities,
while the entropy is concave in all other finite directions.
At all  stationary points of $T$, $\det(H_{ij})$ vanishes.

When some number $m$ 
of the $J^i$ are allowed to become infinite equations (\ref{eq:Jstar}) 
and (\ref{eq:extremumT}) are modified to
\[\J_*^2 = \frac{D-3-2m}{D-3-2m-2n}\]
\[
T_*=\frac{1}{4\pi r_h}\frac{(D-3-2m)(D-3-2m-2n)}{(D-3-2m-n)}\]
respectively.
Indeed in many of the following formulae the thermodynamic properties of a Myers-Perry black hole 
in $D$ dimensions, with $m$ angular momenta
sent to infinity, are seen to be the same as those of a $D-2m$ dimensional black hole 
with all angular momenta finite and the moment of inertia tensor,
which has zero eigenvalues in the infinite directions, suitably
truncated (a caveat to this statement is that we must restrict to $m<\frac{D-3}{4}$,
equation (\ref{eq:Dminus2m})).
This can be seen in the
formulae in the appendix, though $C_J$ and $\det(\Inertia_T)$ are exceptions
and so thermodynamic dimensional reduction using this limit does not work in the canonical ensemble.  
In this sense lower dimensional black holes can be obtained by starting 
from large $D$ and sending more and more of the $J_i$ to infinity.

\subsection{Heat capacities\label{sec:C}}

The heat capacity at constant $J$ is derived in appendix \ref{app:CJ}.
It can be expressed fairly concisely by using the functions
\beq
\Sigma^\pm_n= \sum_{i=1}^\dCSA \frac {\J_i^2}{(1\pm \J_i^2)^n}.\label{eq:Sigmadef}
\eeq
The specific heat at constant $J^i$ is then
\beq
C_J =\frac{4\pi r_h M \X}  
{\Bigl[\X^2-(D-2)\Bigl(\X-4 \Sigma_2^+\Bigr)
\Bigr]},\label{eq:CJ}
\eeq
where 
\beq \X=D-3-2\Sigma_1^+=4\pi r_h T.\label{eq:reducedTdef}\eeq
Equation (\ref{eq:CJ}) generalises the formulae for the specific cases
$D=4$, $5$ and $6$, given in \cite{MPS}, to arbitrary $D$.

As is well known the specific is negative for $J^i=0$,
\beq C_{J=0} = - 4\pi r_h M,\eeq
but can be positive for non-zero $J^i$.

The specific heat at constant $J^i$ (the canonical ensemble) is related to the
specific heat at constant $\Omega_i$ (the grand canonical ensemble) by 
equation (\ref{eq:CJOmegaRelation}).
Alternatively $C_\Omega$ can be evaluated directly for a general $D$
without knowing the moment of inertia explicitly.
The details are left to appendix \ref{app:COmega} and here we just quote the result, 
\beq C_{\Omega} = - \frac{4 \pi r_h M \X \bigl(D - 2 + 2 \Sigma_1^-\bigr)}
{(D - 2) (D-3 + 2\Sigma_1^-)}\,, 
\label{eq:COmega}\eeq
which generalises the $D = 4$ result of \cite{MPS2}   to $D \ge  4.$

To simplify some later formulae It will be convenient to define, in analogy
with (\ref{eq:reducedTdef}),
\[ \overline \X = D-3 +2\sum_{i=1}^\dCSA \frac{\J_i^2}{1-\J_i^2}=D-3 + 2\Sigma_1^-,\]
in terms of which
\[ C_{\Omega} = - \frac{4 \pi r_h M  \bigl(D - 2 + 2 \Sigma_1^-\bigr) \X}{(D - 2)\overline \X}\,, \label{eq:COmegaX}\]

Note the signs: in this notation
\beq \Sigma_1^-(\J^2)=-\Sigma_1^+(-\J^2)\,, \qquad
\overline \X (\J^2)=\X(-j^2) \,.\eeq
There is a curious parallel between the singularities of $\beta C_\Omega$,
where $\overline \X=0$, and extremal black holes for which $t$, and hence $T$,
vanishes.  Since $\overline \X(j^2) = \X(-j^2)$ these are related by mapping
$(J^i)^2 \rightarrow -(J^i)^2$, keeping the entropy constant.

\subsection{Moment of inertia tensor \label{sec:Inertia}}

The isothermal moment of inertia tensor,
\beq
\Inertia_T^{ij}=\left( \frac{\partial J^i}{\partial \Omega_j}\right)_T,\eeq 
is derived in appendix \ref{app:I_T}.
It is 
\beq
\Inertia_T^{ij}= \frac{2 M r_h^2}{D-2}\left\{
\frac{(1+\J_i^2)^2}{(1-\J_i^2)}\delta_{ij} - 
2\J_i \J_j
\left(1+\frac{4}{\overline \X(1-\J_i^2)(1-\J_j^2)}\right)
\right\}.\label{eq:IT}
\eeq
Equation (\ref{eq:IT}) generalises the formulae for the particular cases
$D=4$, $5$ and $6$ derived in \cite{MPS}.

The isentropic moment of inertia tensor
\beq
\Inertia_S^{ij}=\left( \frac{\partial J^i}{\partial \Omega_j}\right)_S\eeq
was given for general $D$ in \cite{C-MP}.  A derivation is outlined 
in appendix \ref{app:I_S} and it has the form 
\beq
\Inertia_S^{ij}= \frac{2 M r_h^2}{(D-2)}
\left\{\frac{(1+\J_i^2)^2}{(1-\J_i^2)}\delta_{ij} - 
\frac{2\J_i \J_j}{\bigr(D-2+2\Sigma_1^-\bigr)}\frac{(1+\J_i^2)(1+\J_j^2)}{(1-\J_i^2)(1-\J_j^2)}
\right\}.\label{eq:IS}
\eeq
The determinant of the isentropic moment of inertia tensor is
\beq
\det{\Inertia_S}=
\left(\frac{2 M r_h^2}{D-2}\right)^\dCSA 
\frac{(D-2)}{(D-2 +2\Sigma_1^-)}
\prod_{i=1}^\dCSA \frac{(1+\J_i^2)^2}{(1-\J_i^2)}. \label{eq:detIS}
\eeq
Note that, in the determinant of the Hessian for the grand canonical ensemble
(\ref{detG}), the factor $D-2 +2\Sigma_1^-$ in the denominator of (\ref{eq:detIS}) 
exactly cancels the same factor in the numerator of (\ref{eq:COmega}).

\subsection{Stability analysis in the canonical and grand canonical ensembles}

 In this section we examine the thermodynamic stability of Myers-Perry black holes in the canonical and the grand canonical ensembles, using the formulae 
of sections \S\ref{sec:C} and \S\ref{sec:Inertia}.  We first summarise the well 
known case of $D=4$ and the results of \cite{MPS}  for $D=5$ and $6$,
before going on to describe  the situation for general $D$.

\subsubsection{D=4\label{sec:D=4}}

The case $D=4$ is well known, but is included here for completeness.
In four-dimensions $\dCSA=1$ and there is only one $J$.
The temperature is
\beq
T=\frac{1}{4\pi r_h} \left(\frac{1-j^2}{1+j^2}\right)
\eeq
so we must restrict to $0\le j \le 1$, with $j=1$ being extremal.
$C_J$ evaluates to
\beq
C_J=-2\pi r_h^2\left(\frac{(1-j^2)(1+j^2)^2}{1-6j^2-3j^4}\right)
\eeq 
which is positive for $\frac{2}{\sqrt{3}} -1 <j^2 < 1$, (in terms of $J$ and $M$, 
\beq \frac{J^2}{M^4}=\frac{4\J^2}{(1+\J^2)^2},\eeq
and these limits corresponding to $2\sqrt{3} -3 < \frac {J^2}{M^4} < 1$).

The isothermal moment of inertia is 
\beq
\Inertia_T :=\frac{r_h^3}{2}(1-6j^2-3j^4),
\eeq
which is only positive when $C_J$ is negative.
Indeed
\beq
\beta C_ J \Inertia_T = -4\pi^2 r_h^6  (1+j^2)^3,\label{eq:4DCJdetIT} 
\eeq
clearly illustrating that Kerr metrics are thermodynamically unstable in the grand canonical ensemble for all values of the angular momentum: when the specific heat is positive the moment of inertia is negative and vice-versa. 
%\cite{TL}).  
Note that the pole in $C_J$ exactly cancels a zero in $\Inertia_T$ --- a phenomenon that we shall see persists for all $D$.

Equation (\ref{eq:equaldets})
immediately shows that an instability must be present
in the canonical ensemble, though
the full story is a little simpler there.
Explicitly
\beq \beta C_\Omega = -8\pi^2 r_h^3,\qquad \Inertia_S = \frac{r_h^3}{2}(1+\J^2)^3,
\eeq
and indeed $\beta C_J \Inertia_T=\beta C_\Omega \Inertia_S$ as it should be,
even though the instability can shift between the specific heat and the moment
of inertia in the former case while it always resides in the specific heat
in the latter, the isentropic moment of inertia always being positive.

\subsubsection{D=5 \label{sec:D=5}}

In five-dimensions the Hawking temperature is
\beq
T= \frac{1}{2\pi r_h}\frac{1-\J_1^2 \J_2^2}{(1+\J_1^2)(1+\J_2^2)},
\eeq
so $\J_1^2 \J_2^2 \le 1$, with the locus of extremal black holes being the hyperbolae
$\J_1^2 \J_2^2 = 1$.

\begin{figure}[!h]
\centerline{\includegraphics[width=7cm]{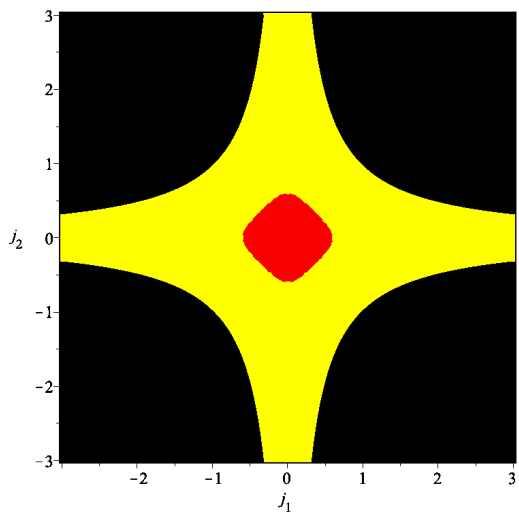}
}
\caption{$C_J$ for 5-dimensions in the $\J_1-\J_2$ plane.
It is negative in the red region, positive in the yellow region and diverges
on the boundary between these two regions. $C_J$ vanishes on the outer hyperbolae, because that is the $T=0$ locus, and is negative in the black region,
where $T<0$.
 (This figure is essentially the same as one in \cite{MPS}, using
slightly different variables.)}
\label{fig:CJ5}
\end{figure}

The specific heat at constant $J$ and the isothermal 
moment of inertia tensor are easily
determined from the general formulae in \S\ref{sec:thermodynamics}, with $\dCSA=2$, but the explicit forms are not illuminating and we shall resort to a graphical representation.
The specific heat is positive in the region of the $\J_1-\J_2$ plane indicated in figure \ref{fig:CJ5}: 
it diverges on the boundary of the red inner region and vanishes on the outer hyperbolae
(the latter being the $T=0$ curve); and is positive in the yellow region 
enclosed by the curves.

\begin{figure}[!h]
\centerline{\includegraphics[width=7cm]{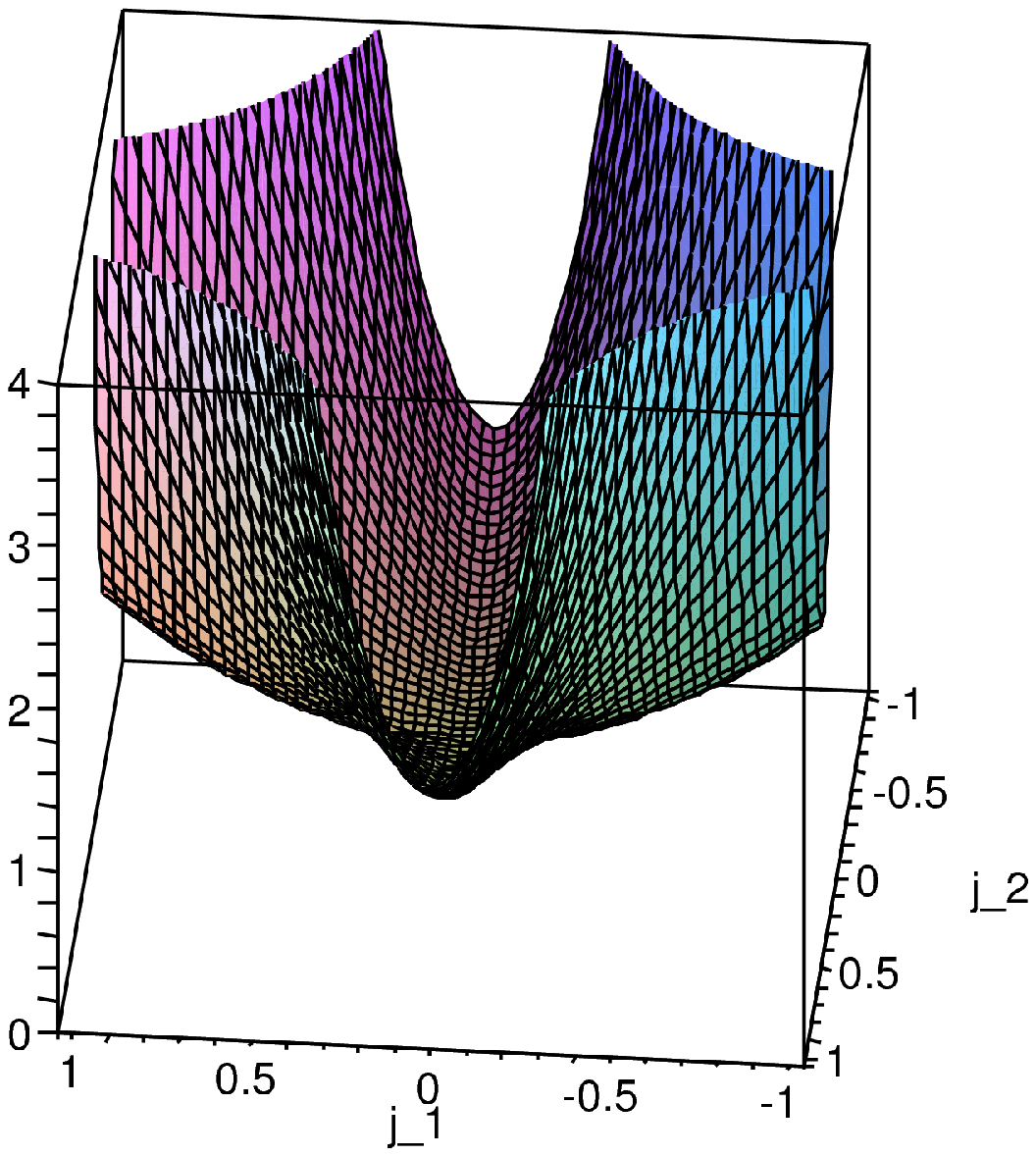}
\includegraphics[width=7cm]{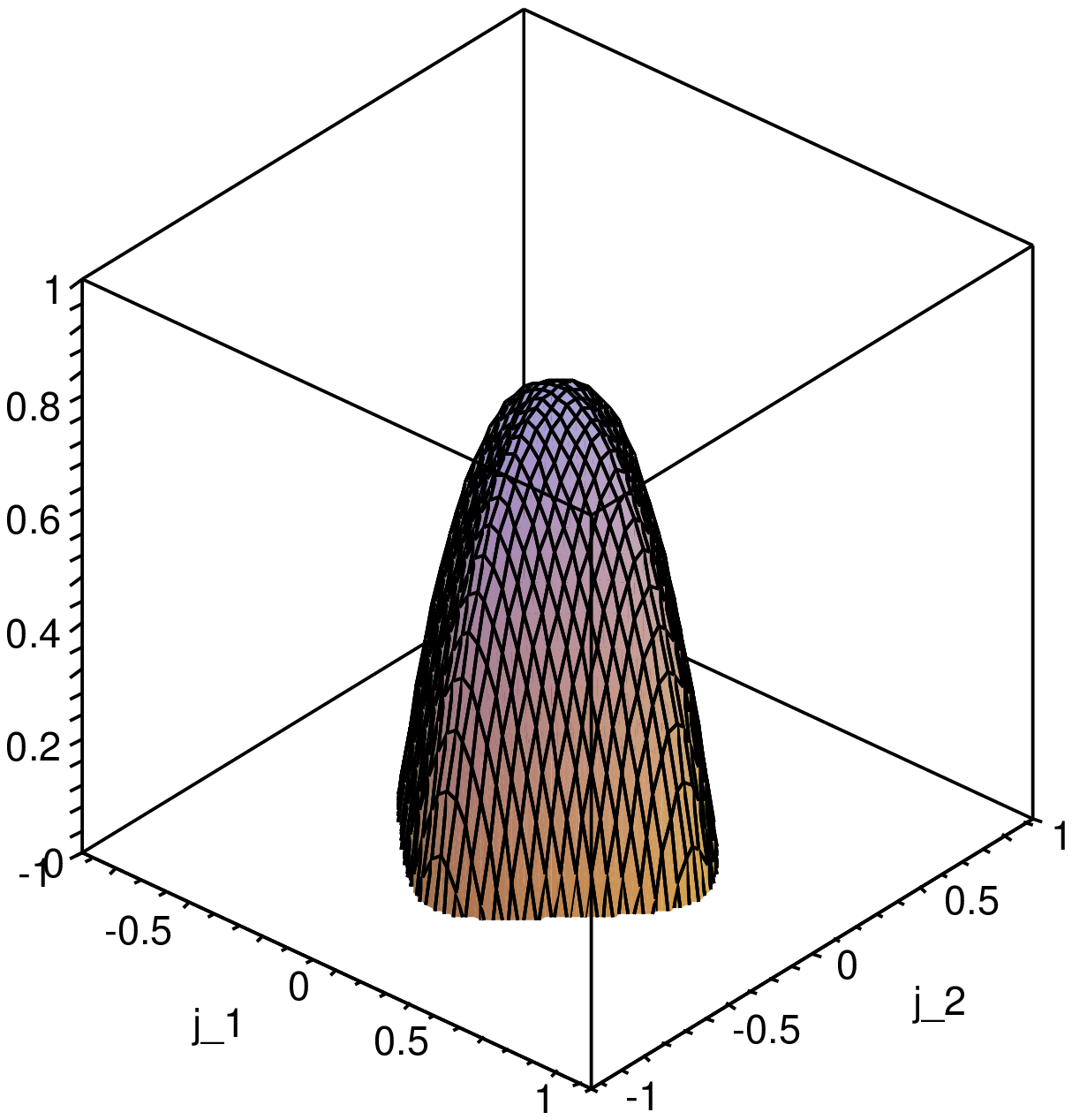}}
\caption{Eigenvalues of the isothermal moment of inertia tensor, $\Inertia_T^{ij}$, for $D=5$. The boundary of the central peak in the right-hand figure, the zero locus of this eigenvalue, coincides with the boundary between the red and
yellow regions
of $C_J$ in figure \ref{fig:CJ5}}
\label{fig:I5}
\end{figure}

The eigenvalues of the isothermal moment of inertia tensor are plotted in figure \ref{fig:I5}. Both eigenvalues are positive for small $\J_i$, and
one is always positive, but the other vanishes on the same curve that bounds 
the red region in figure \ref{fig:CJ5} and is negative outside this region.
The innermost surface on which the moment of inertia
tensor develops a negative eigenvalue is termed the 
{\it ultra-spinning surface} in reference \cite{DFMS} and
it was shown there that there is no ultra-spinning surface in the microcanonical ensemble for a singly spinning Myers-Perry black hole $D=5$.
In contrast we see here that there is an ultra-spinning surface for $\Inertia_T$
in the grand canonical ensemble --- the concept of an ultra-spinning surface depends on the ensemble
used.
  
Thermodynamic instability can nevertheless be seen directly 
from
\beq
C_J \det \Inertia_T = -\frac{3r_h^{11}\pi^4}{32} 
(1+\J_1^2)^4 (1+\J_2^2)^4<0.
\eeq
$\Inertia_T$ has a  negative eigenvalue when $C_J$ is positive and
when $\Inertia_T$ has two positive eigenvalues, $C_J<0$.
Hence $C_J \det \Inertia_T$ is always 
negative for any black-hole, and so these black-holes
are thermodynamically unstable for any choice of angular-momenta with positive temperature.

\subsubsection{D=6 \label{sec:D=6}}

In $6-D$ the temperature 
\beq T=\frac{1}{4\pi r_h} \frac{(3+\J_1^2 + \J_2^2 -\J_1^2\J_2^2)}{(1+\J_1^2)(1+\J_2^2)}\eeq
again vanishes on hyperbolae in the $\J_1$-$\J_2$ plane.

\begin{figure}[!h]
\centerline{\includegraphics[width=7cm]{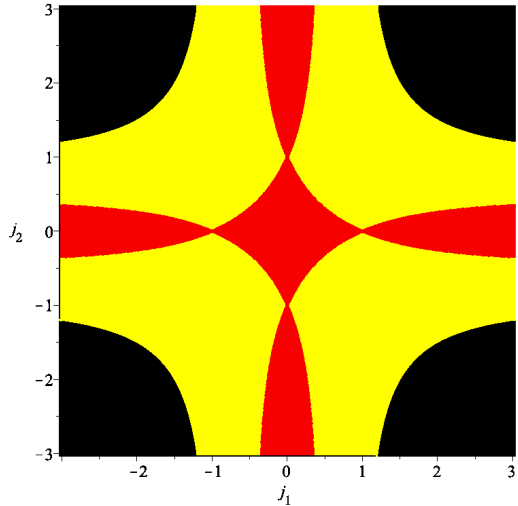}
}
\caption{$C_J$ in 6-dimensions, with the same colour coding as in figure
\ref{fig:CJ5}. 
(This figure is essentially the same as one in \cite{MPS}, using
slightly different variables.)
}
\label{fig:CJ6}
\end{figure}

\begin{figure}[!h]
\centerline{\includegraphics[width=7cm]{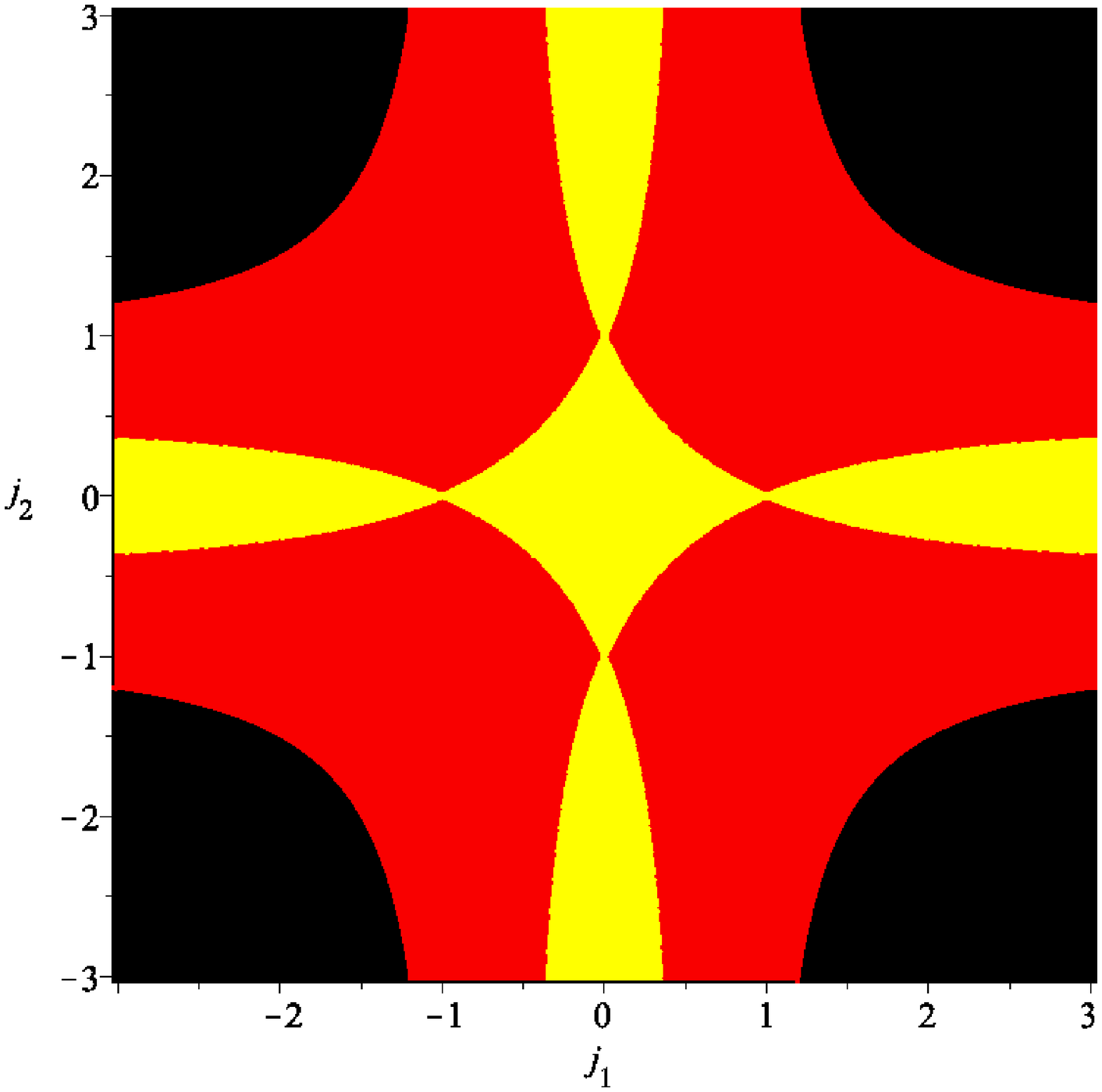}
\includegraphics[width=7.6cm]{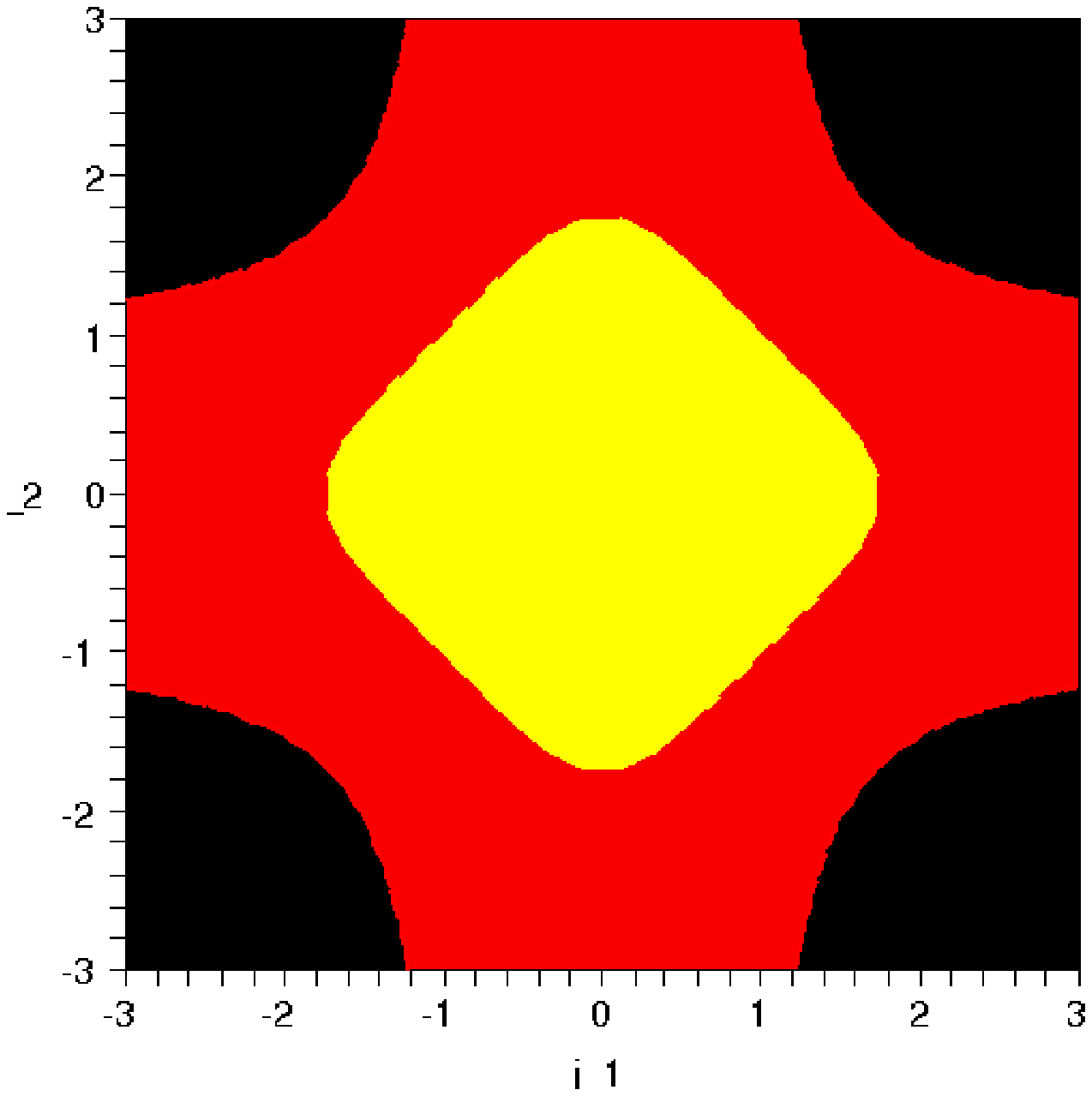}}
\caption{Eigenvalues of the isothermal moment of inertia tensor, $\Inertia_T^{ij}$,
for $D=6$ (same colour coding as in fig \ref{fig:CJ5}).}
\label{fig:I6}
\end{figure}

The specific heat at constant $J$ looks a little more complicated
than in $5D$, but only because some of the hyperbolae overlap.  Figure \ref{fig:CJ6} displays similar information 
to figure \ref{fig:CJ5}, but for $D=6$ ---
this figure is essentially the same as one in \cite{MPS} and is reproduced here for
comparison with figure \ref{fig:I6}.

The regions in the $\J_1$-$\J_2$ plane where the eigenvalues of the isothermal moment of inertia tensor take positive values are plotted in figure \ref{fig:I6}.
The yellow cross-shaped shaded region in the left hand graph of figure,
where one of the eigenvalues of $\Inertia_T$ is positive, exactly co-incides with the inner red region of $C_J$ in figure \ref{fig:CJ6}, thus the product of these two quantities is always negative.
The determinant 
\beq \beta C_J \det\Inertia_T =-\frac {32} {27} \pi^5 r_h^{15}
\left(\frac{(1+\J_1^2)^5(1+\J_2^2)^5}
{3 - \J_1^2 - \J_2^2 - \J_1^2 \J_2^2}\right),
\eeq
is thus negative in the central yellow region of the other eigenvalue of
$\Inertia_T$, indicated in the right hand picture 
in figure \ref{fig:I6}, but it is positive in the red region of the right hand figure,
where both eigenvalues of $\Inertia_T$ are negative and $C_J>0$.
While the determinant itself can be positive there is no regime in which $C_J$ and 
both eigenvalues of $\Inertia_T$ are simultaneously positive. The grand canonical ensemble for six-dimensional Myers-Perry black holes is thus everywhere unstable.

\subsubsection{General $D\ge 7$ \label{sec:anyD}}

In \S\ref{sec:D=4}---\ref{sec:D=6} the stability properties of Myers-Perry
black holes in the canonical ensemble were analysed in terms of $C_J$ and 
isothermal moment of inertia, associated to the canonical ensemble through
(\ref{eq:d2U}).
We focus in this section on the grand canonical ensemble, partly because the
canonical ensemble has already been analysed 
(albeit only for $D=4,5,6$) but primarily because it is algebraically
somewhat simpler than the canonical ensemble.  The general principles
of \S \ref{sec:EnsembleRelation} 
ensure that the stability properties are the same: since $\partial_A\partial_B U$
and $-\partial^A\partial^B G$ are inverses of each other their signature
is the same.

One necessary condition for thermodynamic stability is
\bea \beta C_\Omega = -\frac{16\pi^2 M r_h^2}{(D-2)}
\frac{\bigl(D-2+2\Sigma_1^-\bigr)}{\bigl(D-3+2\Sigma_1^-\bigr)}>0,
\label{eq:betaCOmega}
\eea
in particular $\beta C_\Omega$ is negative for non-rotating black holes
with all $\J_i=0$.
More generally  we must examine the condition 
\beq \frac{\bigl(D-2+2\Sigma_1^-\bigr)}{\bigl(D-3+2\Sigma_1^-\bigr)}<0.
\label{eq:COmegaratio}
\eeq
In terms of the variables
\beq x_i = \frac{\J_i^2}{1-\J_i^2},\eeq 
(\ref{eq:COmegaratio}) is a simple ratio of linear functions
\beq \frac{\bigl(D-2+2\sum_i^\dCSA x_i)}{\bigl(D-3+2\sum_i^\dCSA x_i\bigr)}<0,
\eeq
and positivity of $\beta C_\Omega$
requires that
\beq -\frac{D- 2} {2}<\sum_{i=1}^\dCSA x_i < -\frac{D-3} {2},\label{eq:slab}\eeq
{\it i.e.} the $x$'s are constrained to lie between two hyperplanes
in $x$-space, which never intersect for finite $x_i$. 
However $x_i$ diverges when $j_i^2$ passes through 1, and this
description pushes some subtleties around $\J_i^2=1$ out to infinity.  

\begin{figure}[!h]
\centerline{\includegraphics[width=7cm]{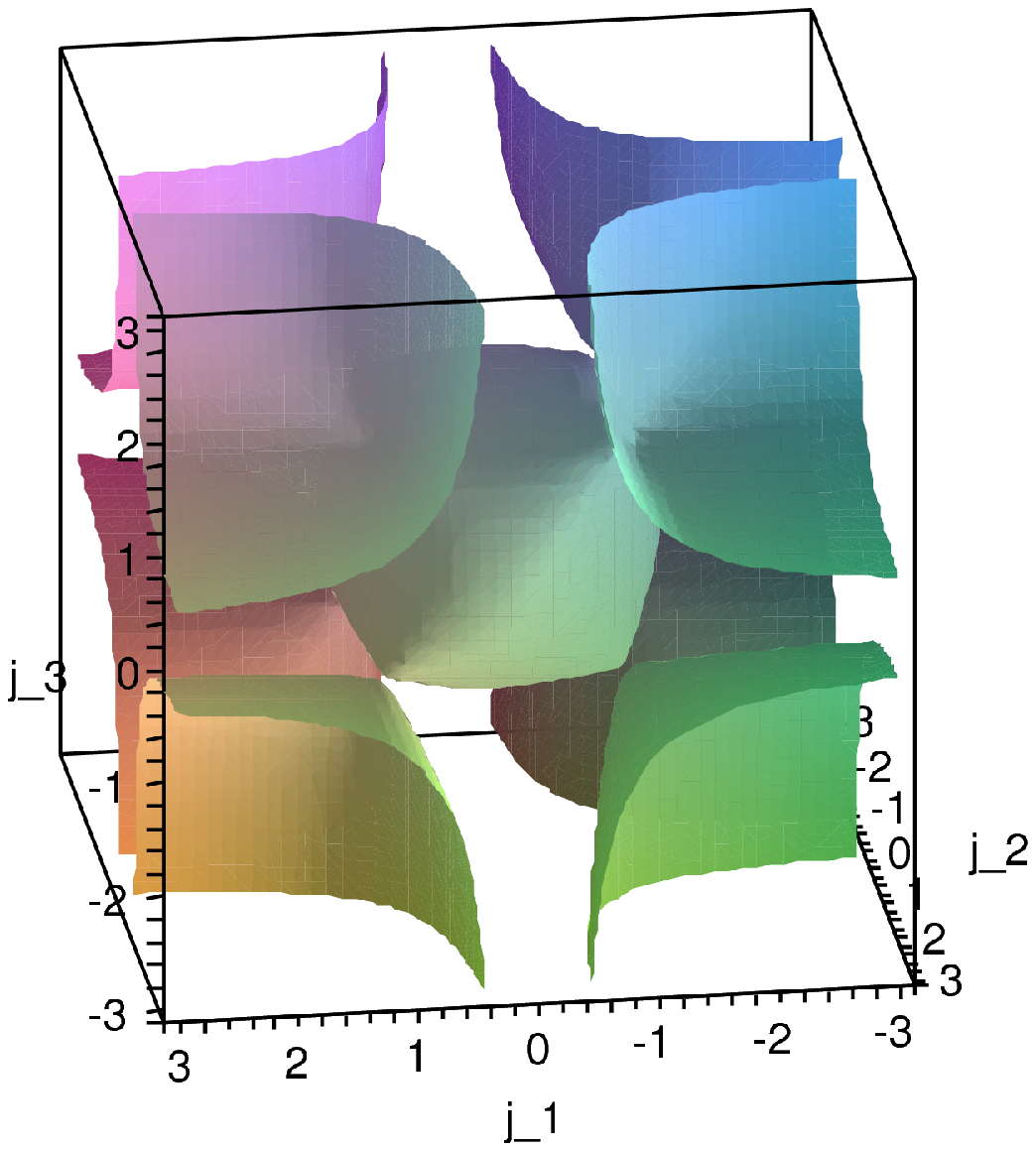}
\includegraphics[width=7cm]{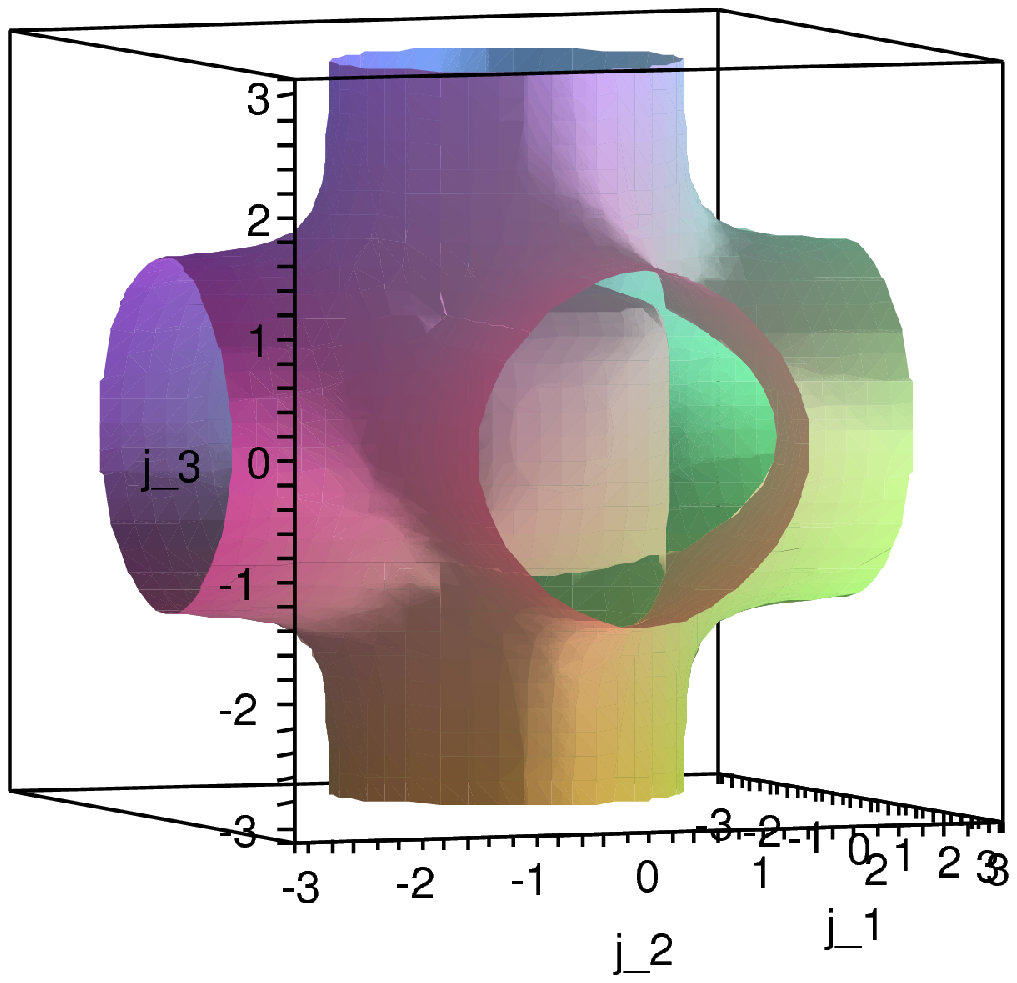}}
\centerline{\includegraphics[width=7cm]{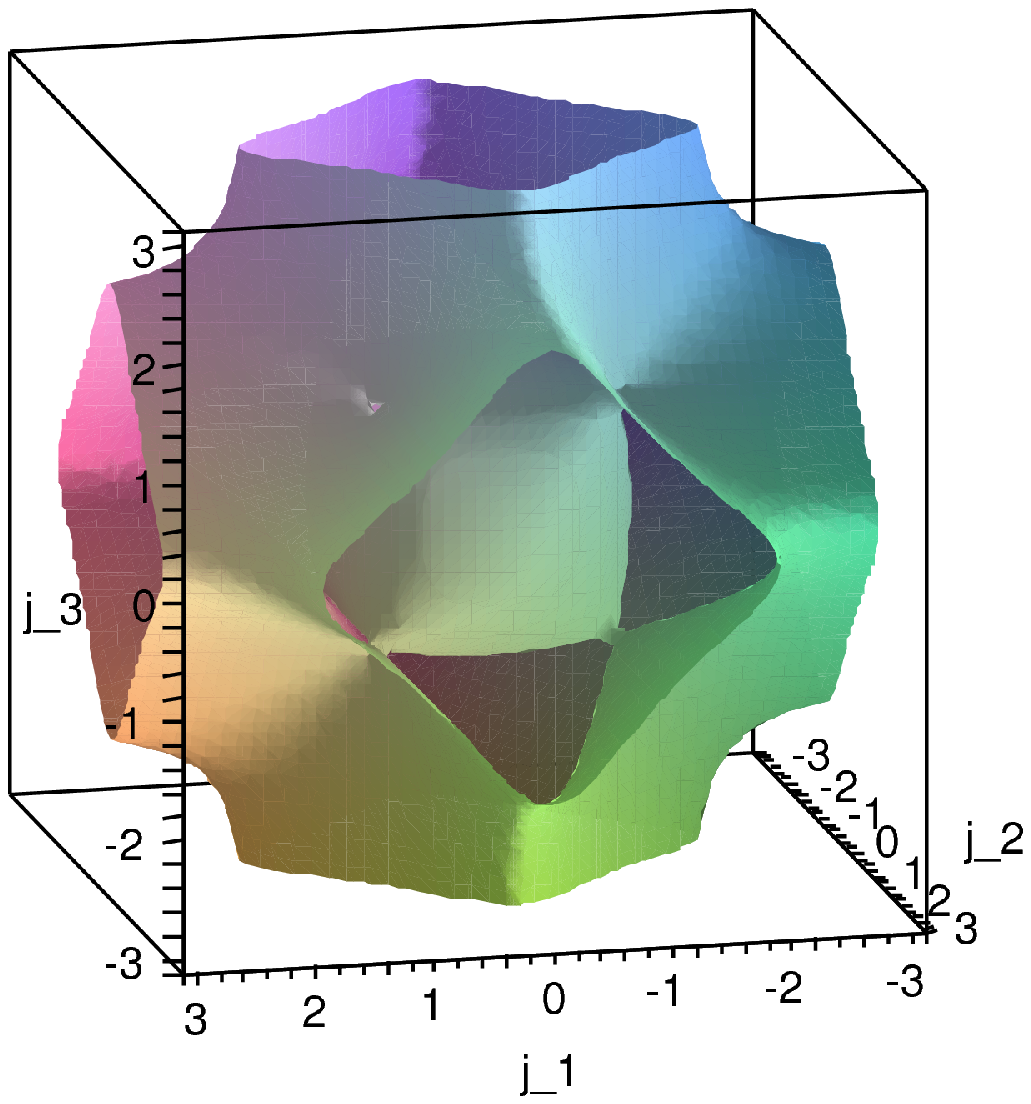}}
\caption{Hypersurfaces on which $\beta C_\Omega$ changes sign in $7$ and $8$ dimensions.
Top-left is  ${\cal C}_{7,3}=0$, on which $\beta C_\Omega$ diverges in
$D=7$; top-right figure is  ${\cal C}_{8,2}=0$, on which $\beta C_\Omega$ vanishes in
$D=8$; the bottom figure is both ${\cal C}_{7,2}=0$
and ${\cal C}_{8,3}=0$, on which $\beta C_\Omega$ vanishes in
$D=7$ and diverges in $D=8$.}
\label{fig:C-hypersurface}
\end{figure}

So we consider instead the condition 
\bea
\frac{\bigl(D-2+2\Sigma_1^-\bigr)\prod_i^\dCSA(1-\J_i^2)}{\bigl(D-3+2\Sigma_1^-\bigr)\prod_i^\dCSA(1-\J_i^2)}<0.
\eea
This ratio can only change sign either across the hypersurface
\beq \bigl(D-2+2\Sigma_1^-\bigr)\prod_i^\dCSA(1-\J_i^2)=0,\label{eq:C_2}
\eeq
where it has a zero, or across the hypersurface
\beq
\bigl(D-3+2\Sigma_1^-\bigr)\prod_i^\dCSA(1-\J_i^2)=0,\label{eq:C_3}
\eeq
where it has a pole.  Both these hypersurfaces are of the form
\beq
{\cal C}_{D,s}:=(D-s)\prod_i^\dCSA(1-\J_i^2) + 2\sum_{k=1}^\dCSA \left(\J_k^2 
\prod_{i\ne k}^\dCSA(1-\J_i^2)\right)=0,\label{eq:C_s}
\eeq
with $s=2$ or $3$.
If any  $\J_i^2=1$, for example if $\J_1^2=1$,
then 
\beq
{\cal C}_{D,s}=2 \prod_{i=2}^\dCSA(1-\J_i^2)=0,
\eeq
and at least one other $\J_i^2$ must be one, the remaining $\J_i$'s,
$\dCSA-2$ of them, are arbitrary.
Indeed the hypersurfaces ${\cal C}_{D,2}$ and  ${\cal C}_{D,3}$
intersect on a manifold of co-dimension two, which is 
actually a flat ${\mathbf R}^{N-2}$ in $\J$-space). 
In $D=7$ for example, the relevant hypersurfaces are ${\cal C}_{7,3}$ and ${\cal C}_{7,2}$, while in $D=8$ they are ${\cal C}_{8,3}$ and ${\cal C}_{8,2}$.
These various hypersurfaces  are shown in figures \ref{fig:C-hypersurface}.

We can determine whether or not $\beta C_\Omega$ changes sign across these
hypersurfaces by following it out along rays from the origin in specific directions. For example in the direction $\J_1=\J$, $\J_2=\cdots=\J_\dCSA=0$,
\beq
\frac{{\cal C}_{D,2}}{{\cal C}_{D,3}}=\frac{D-2-(D-4)\J^2}{D-3-(D-5)\J^2},
\eeq
which is negative between ${\cal C}_{D,2}=0$ and  ${\cal C}_{D,3}=0$
where 
\beq
\frac{D-2}{D-4}< \J^2  < \frac{D-3}{D-5}.
\eeq
Thus $\beta C_\Omega$ does indeed change sign when 
it crosses either of the
hypersurface ${\cal C}_{D,2}=0$ or ${\cal C}_{D,3}=0$ in this direction
(in this specific direction each hypersurface has only one branch, and so is
only crossed once).  
We note in passing that the hypersurface ${\cal C}_{D,3}$, on which
$j^2=\frac{D-3}{D-3-5}$, coincides with the surface on which $T$ is minimised
in the microcanonical ensemble, equation (\ref{eq:Jstar}) with $n=1$.

The determinant of the Hessian for Myers-Perry black holes,
in the grand canonical ensemble,
(\ref{eq:minusd2G}) is derived in appendix \ref{app:I_S}, equation (\ref{app:COmegadetI}),
\beq
\beta C_\Omega \det\Inertia_S = -\frac{8\pi^2(D-2)}{(D-3+2\Sigma_1^-)} \left(\frac{2 M r_h^2}{D-2}\right)^{\dCSA+1}
\prod_{i=1}^\dCSA \frac{(1+\J_i^2)^2}{(1-\J_i^2)}.
\label{eq:COmegadetI}
\eeq
Thus the factor $D-2+\Sigma_1^-$ in the numerator of $C_\Omega$, giving rise
to a zero in the specific heat, is cancelled by a similar factor in the
denominator of $\det(\Inertia_S)$.  
It is actually more convenient to examine ${\Inertia_S}^{-1}$ 
rather than $\Inertia_S$, as it has the sightly simpler form (\ref{app:ISInverse}),
\beq
\bigl(\Inertia_S^{-1}\bigr)_{ij}=\left(\frac{\partial \Omega_i}{\partial J_j}\right)_S = 
 \frac{(D-2)}{2 M r_h^2}
\left\{
\frac{(1-\J_i^2)}{(1+\J_i^2)^2}\delta_{ij}
+\frac{2\J_i \J_j}{(D-2)(1+\J_i^2)(1+\J_j^2)}
 \right\}.\label{eq:ISInverse}
\eeq

Focusing first on the determinant, stability requires
\beq
\det{\Inertia_S}^{-1}=
\left(\frac{D-2}{2 M r_h^2}\right)^\dCSA 
\frac{(D-2 +2\Sigma_1^-)}{(D-2)}
\prod_{i=1}^\dCSA \frac{(1-\J_i^2)}{(1+\J_i^2)^2}>0.\label{eq:detIS+}
\eeq
Of course positivity, while necessary for stability, is not sufficient, (\ref{eq:detIS+}) is satisfied when there are an even number of negative eigenvalues,
but we do know that $\det{\Inertia_S}^{-1}$ can only change sign when ${\cal C}_{D,2}=0$.

To understand the eigenvalue structure in more detail consider first
the two cases $D=7$ and $D=8$. 
The relevant surfaces are ${\cal C}_{7,2}$ and ${\cal C}_{7,2}$ shown in figure \ref{fig:C-hypersurface}.
Each surface ${\cal C}_{D,2}$ consists of two branches, on which at least one eigenvalue 
of (\ref{eq:ISInverse}) must vanish, 
touching at the symmetric point $\J_1^2=\J_2^2=\J_3^2=1$ where two eigenvalues
vanish and the third is positive.  These two surfaces
divide the parameter space into three regions.
All three eigenvalues are positive in the interior region, 
inside the inner surface that is visible through the holes in the outer surface,
because they are positive at the origin where ${\Inertia_S}^{-1}$ is a positive 
multiple of the identity matrix.

We can determine explicitly how many negative eigenvalues there 
are in the intermediate region between the two surfaces simply
by checking the number at any one point in the region, there must be the same number at any other point in the region
as none can change sign unless we cross one of the surfaces.  Similarly we can find the number in the exterior region outside both surfaces.

For the region between the two surfaces we need merely set $\J_1=\J_2=0$ 
and choose $\J_3^2=\J^2$ large enough to ensure that we are outside the
interior region. 
Then $ \left(\frac{2 M r_h^2}{D-2}\right) {\Inertia_S}^{-1}$ in (\ref{eq:ISInverse}) is diagonal and
the eigenvalues are easily read off as
\beq  1, \qquad 1 \qquad 
\hbox{and}  \qquad  
\frac{(D-2)-(D-4)\J^2}{(D-2)(1+\J^2)^2}, \eeq
Hence there is one negative eigenvalue if $\J^2>\frac{D-2}{D-4}$,
with $\J^2=\frac{D-2}{D-4}$ marking the boundary of the interior region in the $\J_3$-direction.

For the region exterior to both the surfaces we can set $\J_1^2=\J_2^2=\J_3^2=\J^2$, with $\J$ large enough to ensure that we are in the exterior region. 
Now (\ref{eq:ISInverse}) shows
that that the eigenvalues of $\left(\frac{r_h S}{2\pi}\right) {\Inertia_S}^{-1}$ are
\beq   \frac{(1-\J^2)}{(1+\J^2)^2}, \qquad 
 \frac{(1-\J^2)}{(1+\J^2)^2} \qquad 
\hbox{and}  \qquad  
 \frac{(D-2)-(D-8)\J^2}{(D-2)(1+\J^2)^2}.\eeq
Hence there are two degenerate negative eigenvalue if $\J^2>1$
and always one other positive eigenvalue for $D=7$ or $8$.
We have thus shown that, for every point in the interior region,
${\Inertia_S}^{-1}$  has three positive eigenvalues, every point in the intermediate region has two positive eigenvalues 
and one negative one while 
every point in the exterior regions has two negative
eigenvalues and one positive one.
Since $\beta C_\Omega$ vanishes on the same surfaces, and is negative
in the interior region since it is negative at the origin,
we see that the canonical ensemble is never stable in $7$ or $8$
dimensions.

The above analysis is easily extended to $D>8$.  We only need determine the 
signs of the eigenvalues
of (\ref{eq:ISInverse}) in special directions $\J_1=\cdots \J_n=\J$,
$\J_{n+1}=\cdots \J_\dCSA=0$, and this gives the signs in each of regions
separated by the roots of 
\beq
{\cal C}_{D,2}=\bigl[D-2 -(D-2-2n)\J^2\bigr](1-\J^2)^{n-1}=0.  
\eeq
The number of regions in any specific direction is determined by the number of roots,
with $\J^2>0$, and the greatest number is when $n=\dCSA$:
there are then $\dCSA-1$ such roots and the different branches
of ${\cal C}_{D,2}=0$ divide $\J$-space into $\dCSA$ regions.

The form of (\ref{eq:ISInverse}) in these directions is
\beq
\left(\frac{2 M r_h^2}{D-2}\right){\Inertia_S}^{-1}=
\begin{pmatrix}
\frac{1-\J^2}{(1+\J^2)^2}{\mathbf 1}_{n\times n} +
\frac{2}{D-2}\frac{\J^2}{(\J^2+1)^2} {\AllOne}_{ n\times n} & 0 \\
0 & {\mathbf 1}_{(N-n)\times (N-n)}
\label{eq:AIplusBQ}
\end{pmatrix},
\eeq
where ${\mathbf 1}_{d\times d}$ are $d\times d$ identity matrices and
${\AllOne}_{ n\times n}$ is the $n\times n$ matrix whose entries are all
one.\footnote{We use the same notation as \cite{DFMS}.}  
There are $N-n$ eigenvalues $+1$ and the remaining eigenvectors ${\mathbf V}=(V_1,\ldots,V_n,0\ldots,0)^t$ and eigenvalues $\lambda$
are determined by
\beq
\frac{1-\J^2}{(1+\J^2)^2}V_i 
+\frac{2}{(D-2)}\frac{\J^2}{(1+\J^2)^2}\sum_{k=1}^n V_k =\lambda V_i.
\eeq
There are two possibilities:
\begin{enumerate}
\item  $\sum_{k=1}^n V_k \ne 0$: this requires $V_1=\cdots =V_n$
which implies that, for $i=1,\ldots,n $,
\beq
\lambda  = \frac{1-\J^2}{(1+\J^2)^2} +  \frac{2 n}{(D-2)} \frac{\J^2}{(1+\J^2)^2}
= \frac{D-2-(D-2-2n)\J^2}{(D-2)(1+\J^2)^2}.\eeq
For $1\le n < \dCSA$, this
configuration returns a negative eigenvalue for $\J^2>\frac{D-2}{D-2-2n}$,
while for  
$n = \dCSA = \frac{D-1-\epsilon}{2}$
this eigenvalue is positive for all values of $\J^2$.

\item $\sum_{k=1}^n V_k=0$: giving $\lambda = \frac{1-\J^2}{(1+\J^2)^2}$,
and 
\beq {\mathbf V}=\begin{pmatrix}\ 1\\-1\\\ 0\\\ 0\\\vdots\\\ 0\end{pmatrix},\quad 
\begin{pmatrix}\ 1\\\ 1\\-2\\\ 0\\\vdots\\0\end{pmatrix},\quad \ldots
\quad\begin{pmatrix}\ 1\\\ 1\\\vdots\\\ 1\\\ 1\\-n\end{pmatrix}.
\eeq 
This requires $n\ge 2$ and has degeneracy $n-1$. This gives $n$ negative 
eigenvalues when $\J^2>1$.
\end{enumerate}

The overall picture is then that there are $N-1$ branches to the 
hypersurface ${\cal C}_{D,2}$ which divide $\J$-space into $\dCSA$ regions.
All eigenvalues of ${\Inertia_S}^{-1}$ are positive at the origin and at every point inside
the first branch.  Every time a branch is crossed by a ray emanating from the origin, one of the positive eigenvalues of ${\Inertia_S}^{-1}$ changes
sign and becomes negative until, in the outer region after all $N-1$ branches
have been crossed, there are $N-1$ negative eigenvalues and one remaining
positive one. The only region in which ${\Inertia_S}^{-1}$, and hence  ${\Inertia_S}$, is a positive matrix is the innermost one.
But we have already seen that  $\beta C_\Omega$ is negative in the 
innermost region, hence the canonical ensemble is always unstable for any choice of metric parameters in any $D$.

In addition to the positive mass Myers-Perry black holes in odd dimensions
there are also negative mass Myers-Perry black holes \cite{MP}.
However, as pointed out in \cite{GK}, there is a subtlety with these
space-times: geodesics are repelled from the would-be event horizon
and do not pass through it, so in a sense there is no event horizon.
Nevertheless one expects a non-zero Hawking temperature, determined
by demanding regularity of the Euclidean time metric, so the entropy
cannot be zero.  The thermodynamics of these space-times is not analysed
here, but would be an interesting future project.

\section{Conclusions\label{sec:conclusions}}

 We have compared the microcanonical, the canonical and the grand canonical ensembles in the thermodynamic description of asymptotically flat
rotating black holes in arbitrary dimensions.
These black holes are always thermodynamically unstable
but the thermodynamic instability manifests itself differently in the different
ensembles.  There is however an elegant and 
simple relation between the specific heats and moment
of inertia tensors in the canonical and the grand canonical ensembles,
given by equation (\ref{eq:detITdetIS}),
\beq
C_J \det\Inertia_T = C_\Omega \det\Inertia_S.
\eeq

The case of Myers-Perry black-holes has been analysed in detail and 
all extrema of the temperature in the microcanonical ensemble have been found and classified and shown to correspond to inflection points of the entropy.

In the canonical ensemble it has been
shown that, in $D$ dimensions, the specific heat $C_J$ in
equation (\ref{eq:CJ}) vanishes when $T=0$ and changes sign
on a hypersurface in angular momentum space given by
\beq (4\pi r_h T)^2 = (D-2)\left(4\pi r_h T -4 \sum_{i=1}^\dCSA \frac{\J_i^2}{(1+\J_i^2)^2}\right)\label{eq:CJpole}
\eeq
(where $\J_i=\frac {2 \pi J^i}{S}$), on which it diverges.
In the determinant of the Hessian this
singularity in $C_J$ 
is exactly cancelled by an equivalent zero in
$\det(\Inertia_T)$.
There are also singularities in $\det(\Inertia_T)$ when
${\cal C}_{D,3}$ in equation (\ref{eq:C_3}) vanishes.

In the grand canonical ensemble
$C_\Omega$ in equation (\ref{eq:COmega}) also vanishes when $T=0$ and has 
divergences, this time 
on the hypersurface defined by ${\cal C}_{D,3}=0$
rather than that given by (\ref{eq:CJpole}).
In addition $C_\Omega$ also has zeros on the hypersurface  ${\cal C}_{D,2}=0$ in equation (\ref{eq:C_2}).
In the determinant of the Hessian for 
the grand canonical ensemble (\ref{eq:minusd2G})
the zeros of $C_\Omega$ are cancelled by corresponding poles in $\det(\Inertia_S)$ on ${\cal C}_{D,2}=0$. The locus of these
singular points of $\det(\Inertia_S)$ corresponds to a branched hypersurface in angular momentum space which divides the space into $\dCSA$ separate regions.
Every time a branch of this hypersurface is crossed an eigenvalue of
$\Inertia_S$ changes sign and the moment of inertia tensor 
has different signature in the $\dCSA$ separate regions. 
Only the region surrounding the origin in angular momentum space gives a positive definite moment of inertia tensor and this region
corresponds precisely to the region where $C_\Omega$ is negative.

There is a curious relation between the hypersurface ${\cal C}_{D,3}=0$ 
on which both $C_\Omega$ and $\det(\Inertia_T)$ diverge on the one hand
and extremal $T=0$ Myers-Perry black holes on the other:
the algebraic equations defining
these two hypersurfaces are related by analytic continuation $(J^i)^2 \rightarrow -(J^i)^2$, with the entropy held constant.

Our analysis has also shown that, in the microcanonical and the grand canonical ensembles, many of the thermodynamic properties of Myers-Perry
black holes in $D-2$ dimensions can be obtained from those of a black hole
in $D$ dimensions by letting one of the angular momenta in $D$ dimensions 
tend to infinity, keeping the entropy constant.  

The thermodynamic instabilities of Myers-Perry black holes thus have a very rich
structure, beyond that of the ultra-spinning surface upon which the
moment of inertia tensor develops its first negative eigenvalue. 

An obvious direction for future work on this topic is to include a charge on
the black hole and to introduce a cosmological constant to encompass the case of asymptotically anti-de Sitter rotating black holes.  The latter should prove particularly
interesting as the black holes will become thermodynamically stable
when the magnitude
of the cosmological constant is large enough and much could be learned by mapping out the
boundary of the stability region.

\begin{appendix}

\section{Temperature extrema and inflection points of the entropy \label{app:T_ext}}
In this appendix we extend the study in \cite{DFMR} and \cite{DFMS}
to find all isenthalpic ({\it i.e.} constant mass) extrema of $T$
for Myers-Perry black holes, as the $J^i$ are varied 
in asymptotically flat space-times.

At constant $M$ equation (\ref{eq:MJdef}) implies that
\beq 
\left. \frac{\partial a_i}{\partial J^j}\right|_M 
= \frac{(D-2)}{2M}\delta_{ij}\,,
\label{eq:dadJ}
\eeq
with, which the expression
\beq
M=\frac{(D-2)\varpi}{16 \pi} r_h^{D-3} 
\prod_{i=1}^\dCSA\left(1+\J_i^2\right)
\eeq
for the mass, gives
\beq 
d M=0\qquad\Rightarrow\qquad 
\left.\frac{\partial r_h}{\partial J^i}\right|_M = -\frac{(D-2)}{4\pi T M}\Omega_i =  -\frac{(D-2)}{\X M}\,\omega_i\,,\label{eq:drhdJ}
\eeq
where we have defined
\beq \X:=4\pi r_h T =
\left(D-3-2\sum_{i=1}^\dCSA\frac{\J_i^2}{1+\J_i^2}\right)
\quad\hbox{and}\quad
\omega_i=r_h \Omega_i=\frac{\J_i}{1+\J_i^2}
\eeq 
(both $t$ and $\omega_i$ are invariant under $\J_i \rightarrow \frac{1}{\J_i}$).

Equations (\ref{eq:dadJ}) and (\ref{eq:drhdJ}) together imply
\beq
\left.\frac{\partial \J_i}{\partial J^j}\right|_M
= \frac{(D-2)}{2 r_h M}\left(\delta_{ij} +\frac{2}{\X}\frac{\J_i \J_j}{(1+\J_j^2
)} \right).
\eeq
We now have all the information we need to calculate 
$\left.\frac{\partial T}{\partial J^i}\right|_M$
from 
\beq
T=\frac{1}{4\pi r_h}
\left(D-3-2\sum_{i=1}^\dCSA\frac{\J_i^2}{1+\J_i^2}\right),\label{eq:T-extrema-2}
\eeq
we find
\beq
\left.\frac{\partial T}{\partial J^i}\right|_M=
\frac{(D-2)}{4\pi r_h^2 M} \left(1-\frac{2}{1+\J_i^2} - \frac{4\,\Omega^2}{\X} \right)\omega_i\,,
\eeq
where $\Omega^2:=\sum_k \omega_k^2$.

For fixed $M$ extrema of $T$ occur for 
\beq \omega_i =0 \ \Leftrightarrow \ \J_i=0
\qquad\hbox{or}\qquad 1-\frac{2}{1+\J_i^2} - \frac{4\,\Omega^2}{\X}=0.
\label{eq:jstar-condition}\eeq
In particular any finite non-zero $\J_i$ are all equal at an extremum. 
 
It is also possible that some of the $\J_i$ might tend to infinity.
Suppose $m$ of the $\J_i$ diverge as $\J_i\approx \Lambda\rightarrow\infty$.
Then, at fixed finite mass, 
\beq r_h \approx \Lambda^{-\frac{2m}{D-3}},
\eeq 
and hence
\beq
\left.\frac{\partial T}{\partial J^i}\right|_M \approx 
\Lambda^{\frac{4m}{D-3}}\left(1-\frac{2}{1+\J_i^2} - \frac{4\,\Omega^2}{\X} \right)\omega_i\,.\label{eq:extreme-dT}
\eeq
For $\J_i\approx\Lambda$, $\omega_i\approx \Lambda^{-1}$ so
\beq
\left.\frac{\partial T}{\partial J^i}\right|_M \approx 
\Lambda^{\frac{4m}{D-3}-1},\label{eq:Dminus2m}
\eeq
which tends to zero for $\Lambda\rightarrow\infty$ provided $m<\frac{D-3}4$,
which is possible for $D\ge 8$. 

To keep the discussion general we shall suppose 
$n$ of the $\J_i$ are finite and equal, $m$ are infinite  and $\dCSA -n-m$ 
are zero.
Up to permutations of the $\J_i$, extrema of $T$ can only occur for configurations with
\bea 
\J_1=\cdots=\J_n=\J>0,&&
\qquad \J_{n+1}=\cdots = \J_{\dCSA-m}=0,\nonumber\\
&& \kern -60pt 
 \J_{\dCSA-m+1}=\cdots =\J_{\dCSA}=\Lambda\rightarrow\infty\,\label{eq:extrema}
\eea
so we focus on the symmetric angular momentum configurations
\beq
\vec{\J}= \lim_{\Lambda\rightarrow\infty}
(\overbrace{\J,\ldots,\J}^{\vbox{\hbox{$\hskip 8pt \scriptstyle n$}\vskip -5pt \hbox{\small times}}},
\overbrace{0,\ldots,0}^{\vtop{\hbox{$\scriptstyle \dCSA-m-n$}\vskip -5pt \hbox{\small \ times}}},
\overbrace{\vline height 8pt width 0pt
\Lambda,\ldots,\Lambda}^{\vbox{\hbox{$\hskip 8pt \scriptstyle m$}\vskip -5pt \hbox{\small times}}}
).\label{eq:nsymmetricj}
\eeq
The extrema require  $\J$ to satisfy the second equation in (\ref{eq:jstar-condition}), $\J=\J_*$ with
\beq
\J_*^2=\frac{\X+4\Omega^2}{\X-4\Omega^2}.\label{eq:jstar-equation-2}
\eeq
At $\vec{\J}_*$
\beq
\Omega^2= \frac{n\J_*^2}{(1+\J_*^2)^2} \qquad\hbox{and}\qquad
\X=D-3 -2m -\frac{2n\J_*^2}{(1+\J_*^2)}
\eeq
which gives the solution of (\ref{eq:jstar-equation-2}) to be
\beq \J_*^2=\frac{D-3-2m}{D-3-2m-2n}.\label{eq:jstar-extrema}
\eeq

The temperature at these extrema is, from (\ref{eq:T-extrema-2}),
\beq
T_*=\frac{1}{4\pi r_h}\frac{(D-3-2m)(D-3-2m-2n)}{(D-3-2m-n)}.
\eeq
Demanding $T_* \ge 0$ imposes the restriction
\beq
m+n \le \frac{D-3}{2}.\label{eq:mnrange}
\eeq

More generally when the angular momenta are of the form (\ref{eq:nsymmetricj}), but not necessarily at $j_*$,
the temperature is
\beq
T=\frac{1}{4\pi r_h}\frac{(D-3-2m) +(D-3-2m-2n)\J^2}{(1+\J^2)}
\eeq
and vanishes for
\beq \J^2=\J_0^2=\frac{D-3-2m}{2m+2n-(D-3)},\eeq
which is only possible for $m+n \ge \frac{D-3} {2}$.

To analyse the nature of the extrema we need the second derivative of $T$.
A straightforward but tedious calculation gives
\bea
\left.\frac{\partial^2 T}{\partial J^i\partial J^j}\right|_M
&=&\frac{(D-2)^2}{4\pi M^2 r_h^3}\left\{
\left[ \frac{4 \J_i^2}{(1+ \J_i^2)^2} -
 \frac{2\,\Omega^2}{\X}  \frac{(1-\J_i^2)}{(1+\J_i^2)} 
-\frac 1 2
\right]\frac{\delta_{ij}}{1+\J_i^2}\right. \nonumber\\
&&\kern -30pt -\left[ \frac 4 \X \left( \frac{1-\J_i^2}{(1+\J_i^2)^2}
+ \frac{1-\J_j}{(1+\J_j)^2}- \frac 1 4 \right) 
+\frac{8\, \Omega^2}{\X^2}\left(\frac 1 {1+\J_i^2}+ \frac 1 {1+\J_j} + \frac 1 2 \right) 
\right.
\nonumber\\
&&\hskip 70pt\left.\left.+\frac{8}{\X^2}\left(\frac{2\,\Omega^4}{\X} +\widetilde\Sigma  \right)  \right]
\omega_i\omega_j \right\}\,,\label{eq:d2T}\eea 
where
\beq
\widetilde\Sigma:=\sum_k \frac{(1-\J_k^2)\J_k^2}{(1+\J_k^2)^3}.
\eeq
Sums like $\widetilde\Sigma$ crop up frequently in this analysis and
it will prove convenient to define
\beq \Sigma_n^\pm=\sum_{k=1}^\dCSA \frac{\J_k^2}{(1\pm \J_k^2)^n}\eeq
in terms of which 
\beq\widetilde \Sigma =2\Sigma_3^+ - \Sigma_2^+.\eeq

We wish to determine the signs of the eigenvalues of (\ref{eq:d2T}) at $\vec{\J}_*$. There are three cases to consider:
\begin{itemize}

\item At an extremum of the form (\ref{eq:extrema}), if either of the indices $i$ and $j$ is in the range $[\dCSA-n-m,\cdots,\dCSA-m]$
then $\left.\frac{\partial^2 T}{\partial J^i\partial J^j}\right|_{\ \mathbf{j}_*}=0
$
unless the other index is in the same range.
When both $i$ and $j$ are in the range $[\dCSA-n-m,\cdots,\dCSA-m]$,
\bea
\left.\frac{\partial^2 T}{\partial J^i\partial J^j}\right|_{\ \mathbf{j}_*}
&=&-\frac{(D-2)^2}{4\pi M^2 r_{h,*}^3}
\left.\left(\frac{2\,\Omega^2}{\X}+\frac 1 2 \right)
\right|_{\mathbf{j}_*}\delta_{ij}\nonumber\\
&=&-\frac{(D-2)^2}{4\pi M^2 r_{h,*}^3}\frac{(D-3-2m-2n)}{2(D-3-2m-n)}\delta_{ij}\,.
\label{eq:Tmax}\eea
The eigenvalues are all negative in these directions, 
corresponding to a maximum of $T$ around $J_*$.

\item If one of the indices $i$ or $j$ is in the range $[\dCSA-m+1,\ldots,\dCSA]$
and the other is in the range $[1,\ldots,\dCSA-m]$ then 
$\left.\frac{\partial^2 T}{\partial J^i\partial J^j}\right|_{\ \mathbf{j}_*}=0$.
If both are in the range  $[\dCSA-m+1,\ldots,\dCSA]$ then
\beq
\left.\frac{\partial^2 T}{\partial J^i\partial J^j}\right|_{\ \mathbf{j}_*}
\approx \frac{1}{r_{h,*}^3}\frac{1}{\Lambda^2}\approx \Lambda^{\frac{6m}{D-3}-2}\ \mathop{\longrightarrow}_{\Lambda\rightarrow \infty}\ 0\,
\eeq
since $m<\frac{D-3}4$.

\item  If both indices $i$ and $j$ are in the range $[1,n]$,
then the nature of the extremum is determined by 
\beq
\left.\frac{\partial^2 T}{\partial J^i\partial J^j}\right|_{\ \mathbf{j}_*}
=\frac{(D-2)^2}{4\pi M^2 r_{h,*}^3}\,
\bigl(A_* \delta_{ij} + B_* Q_{ij}\bigl)\,,\label{eq:d2Tn}
\eeq
where $Q_{ij}$ is the $n\times n$ matrix whose entries are all unity
and $A_*$ and $B_*$ are ratios of polynomials in $\J_*$.
Since $Q_{ij}$  has $n-1$ zero eigenvalues and one eigenvalue equal to $n$,
(\ref{eq:d2Tn}) has $n-1$ degenerate eigenvalues  $\lambda_1=A_*$ and
one eigenvalue $\lambda_2=A_*+n B_*.$ Evaluating $A_*$ and $B_*$ gives
% T_ij_eigenvalues.mw
\bea
\lambda_1&=&\frac{(D-3-2m)(D-3-2m-2n)^2}{4(D-3-2m-n)^3}>0,\\
\lambda_2&=&\frac 1 4.
\eea
\end{itemize}
Thus $\vec J_*$ is in general a saddle point, with $T$ minimised
in the directions $i=1,\ldots,n$ and maximised
in the directions $i=\dCSA-n-m,\ldots,\dCSA-m$.

A necessary condition for stability is
that the eigenvalues of 
\beq H_{ij}=-\left.\frac{\partial^2 S}{\partial J^i \partial J_j}\right|_M
\label{eq:d2S}
\eeq be positive, \cite{DFMR}.  
Using (\ref{eq:drhdJ}) and
\beq
S=\frac{\varpi}{4}r_h^{D-2}\prod_k(1+\J_k^2),
\eeq 
gives
 \beq
\left.\frac{\partial S}{\partial J^i}\right|_M
=-\frac{2}{\X}\frac{\J_i}{(1+\J_i^2)}.
\eeq
The entropy is thus a monotonically decreasing function of each of the $\J_i$
as $|\J_i|$ increases.
It was shown in \cite{DFMS} that the extrema of the temperature in the two
special cases $n=1$ and $n=\dCSA$ (with $m=0$)
are inflection points of the entropy.
We now show that this is a general property of all the extrema of $T$.

An inflection point corresponds to a zero eigenvalue of the matrix (\ref{eq:d2S}), so we first determine
\bea 
H_{ij}&=& -\left.\frac{\partial^2 S}{\partial J^i \partial J_j}\right|_M \\
&=&\frac{(D-2)} {2 r_h^2 T M} \left\{
\frac{1-\J_i^2}{(1+\J_i^2)^2}\delta_{ij}\right.
 \left.+ \frac{4}{\X}
\left(\frac {1}{1+\J_i^2}  + \frac {1}{1+\J_j^2}  -\frac 1 2 
+ \frac{2}{\X}\Omega^2 \right)
\omega_i \omega_j
\right\}\,,\nonumber
\eea
which agrees with equation (2.13) in \cite{DFMS}, apart from some minor typos.

The eigenvalues at the symmetric points (\ref{eq:nsymmetricj})
are easily determined: 

\begin{itemize}

\item If both indices $i$ and $j$ are in the range $[1,\ldots,n]$ we write
\beq
H_{ij}=\frac{(D-2)} {2 r_h^2  M T} \Bigl({\cal A}\delta_{ij} +{\cal B}Q_{ij} \Bigr)
\eeq
and the eigenvalues of ${\cal A}\delta_{ij} +{\cal B}Q_{ij}$, with
$i,j=1,\ldots,n$, are
\beq \lambda_1={\cal A} \qquad \hbox{and} \qquad \lambda_2 ={\cal A} + n{\cal B},\eeq
where $\lambda_1$ has degeneracy $n-1$. 
These are 
\bea \label{eq:Hlambda}
\lambda_1&=&\frac{(1-\J^2)}{(1+\J^2)^2},\\
\lambda_2 &=& \frac{(D-3-2m)\bigl\{D-3-2m-(D-3-2m-2n)\J^2\bigr\}}
{\bigl\{D-3-2m+(D-3-2m-2n)\J^2\bigr\}^2}\,. \nonumber
\eea

\item If either of the indices $i$ and $j$ is in the range $[n+1,\cdots,N-m]$,
then $H_{ij}$ vanishes unless the other index is in the same range in which case
$\J_i=\J_j=0$  and
\beq
H_{ij}= \frac{(D-2)} {2 r_h^2 T M}\delta_{ij}\,.
\eeq

\item If either of the indices $i$ and $j$ is in the range $(\dCSA-m+1,\ldots,\dCSA)$ then $\lambda=0$.

\end{itemize}

As observed in \cite{DFMS} the eigenvalues are all positive near $J^i =0$ and the
first negative eigenvalue is encountered for $n=1$ when $\J=\frac{D-3}{D-5}$,
which is precisely the inflection point of \cite{EM} at the first temperature
minimum. This hypersurface  $\J=\frac{D-3}{D-5}$ on which $H_{ij}$ first
develops a zero eigenvalue is the ultra-spinning surface of
\cite{DFMS}.  For $D=5$ the ultra-spinning surface in the microcanonical 
ensemble is not closed and $\J_1$ can reach infinity when $\J_2=0$, 
It is shown in \S\ref{sec:D=5}
that the $D=5$ ultra-spinning surface in the canonical ensemble
is closed, see figure \ref{fig:I5}.

At temperature minima, where $\J=\J_*$, the eigenvalues (\ref{eq:Hlambda})
above evaluate to
% H_ij_eigenvalues.mw
\bea \label{eq:Hijeigenvalues}
\lambda_1\bigr|_{\J_*}&=& -\frac n 2 \frac{(D-3-2m-2n)}{(D-3-2m-n)^2}\le 0,
\qquad (n-1)\ \hbox{times};\nonumber\\
\lambda_2\bigr|_{\J_*}&=&0.
\eea
In particular there is always one zero eigenvalue, corresponding to an inflection point in the entropy in the direction of the associated eigenvector..

\section{Specific heat at constant angular momentum \label{app:CJ}}

To calculate the heat capacity at constant $J$
\beq C_J=\left. \frac {\partial M}{\partial T} \right|_J, \eeq
we first observe that (\ref{eq:MJdef}) gives
\beq J^i=\frac {2 M}{D-2} a_i,\eeq
so $J^i= const$ implies 
\beq \frac{d a_i \bigr|_J}{a_i} = - \frac {dM \bigr|_J}{M}. \eeq
Next, combining (\ref{eq:mudef}), (\ref{eq:Sdef}) and (\ref{eq:MJdef}),
$\J_i = \frac{a_i}{r_h}=\frac{2\pi J^i}{S}$
vary as
\beq \left.\frac {d \J_i}{\J_i}\right|_J = -\left.\frac {dS} {S} \right|_J=
\left.\left( \frac{d a_i}{a_i} - \frac{d r_h}{r_h} \right)\right|_J = -\left.\left(\frac {d M}{M} + \frac {d r_h}{r_h}\right)\right|_J\,.
\label{eq:dS1}
\eeq
But explicitly from (\ref{eq:Sdef})
\bea \left.\frac {dS} {S} \right|_J &=& 
\left.(D-2) \frac {d r_h}{r_h}\right|_J 
+ \left.2 \sum_{k=1}^\dCSA \frac{\J_k d \J_k}{1+\J_k^2}\right|_J
\nonumber \\
&=&
\left.(D-2) \frac {d r_h}{r_h}\right|_J 
- 2 \left.\left(\frac {d M}{M} + \frac{d r_h}{r_h} \right)\right|_J
\sum_k^\dCSA \frac{\J_k^2}{1+\J_k^2}.
\label{eq:dS2}
\eea
Equations (\ref{eq:dS1}) and (\ref{eq:dS2}) together now give
\beq
\left. \frac{\partial M}{\partial r_h}\right|_J = \left(\frac \X {D-2-\X} \right) \frac{M}{r_h}\,.\label{eq:dMrh}
\eeq

Similar manipulations on $T=\frac{\X}{4\pi r_h}$ yield
\beq 
\left.\frac{\partial T}{\partial r_h}\right|_J = 
\frac {1} {4\pi r_h^2}
\left\{
\frac{\X^2-(D-2)\left(\X-4 \Sigma_2^+\right)}{D-2-\X}\right\},\label{eq:dTrh}
\eeq
and then (\ref{eq:dMrh}) and (\ref{eq:dTrh}) can be combined to give the specific heat in the canonical ensemble with fixed $J$,

\beq
C_J =\frac{16\pi^2 r^2_h M T}  
{\bigl[\X^2-(D-2)\left(\X-4 \Sigma_2^+\right)
\bigr]}.\label{eq:appCJ}
\eeq
This can be expressed in terms of $M$ and $\J_i$ by noting that
\beq M = \left( \frac{(D-2)\varpi}{16 \pi }\right) r_h^{D-3} \prod_{i=1}^\dCSA \bigl( 1+ \J_i^2 \bigr).\eeq
%giving
%\beq
%C_J = 2\sqrt{\pi} \left\{ \frac {\Gamma\left( \frac {D-1}{2} \right) M^{D-2} }
%{(D-2) \prod_i^\dCSA (1+\J_i^2)} \right\}^{\frac 1 {D-3}}
%\frac{\X}  
%{\left\{(D-2)\left(4 \Sigma_2 -\X\right)
%+\X^2\right\}}.\label{eq:CJ}
%\eeq
%\beq
%C_J = 4\pi \left( \frac {16\pi M^{D-2}}{ (D-2)\varpi}\right)^{\frac 1 {D-3}} 
%\frac{\X}  
%{\Bigl\{ \prod_i^\dCSA (1+\J_i^2) \Bigr\}^{\frac 1 {D-3}}\left\{(D-2)\left(4\sum_i^\dCSA \frac{\J_i^2}{(1+\J_i^2)^2} -\X\right)
%+\X^2\right\}}.
%\eeq
\section{Specific heat at constant angular velocity\label{app:COmega}}

The specific heat at constant $\Omega$ is straightforward to determine, using similar techniques to those of \S\ref{sec:C}.  
In terms of $\J_i$ the entropy (\ref{eq:Sdef}) is
\beq
S=\frac{\varpi r_h^{D-2}}{4}\prod_{i=1}^\dCSA(1+\J_i^2),
\label{eq:lambdaSdef}\eeq
\beq
T=\frac{\X}{4\pi r_h}=\frac{(D-3-2 \Sigma_1^+)}{4\pi r_h}
\label{eq:lambdaTdef}
\eeq
%\beq  
%M=\frac{(1+\lambda r_h^2)\varpi}{16\pi r_h^{2-\epsilon}}
%\prod_{i=1}^\dCSA \frac{r_h^2 + a_i^2}{\\Xi_i}
%\left( D-2 +2\lambda \sum_{i=1}^\dCSA\frac{a_i^2}{\\Xi_i^2}\right) \,,\label{eq:lambdaMJdef}\eeq
and
\beq \Omega_i=\frac{\J_i}{r_h(1+\J_i^2)}.\label{eq:lambdaOmegadef}\eeq

The specific heat at constant angular velocity is defined as
\beq C_\Omega = T\left( \frac{\partial S}{\partial T}\right)_\Omega. \eeq
From (\ref{eq:lambdaOmegadef}) 
\beq \Omega_i=const \qquad\Rightarrow \qquad 
\left.d\J_i\right|_\Omega = \left.
\left(\frac{1+\J_i^2}{1-\J_i^2}\right)\J_i \frac{d r_h}{r_h}\right|_\Omega. \eeq

Using these it is straightforward to show that
\beq
\left. \frac{\partial T}{\partial r_h}\right|_\Omega = -\frac{(D-3 + 2\Sigma_1^-)}{4\pi r_h^2},
\eeq
and
\beq
\left. \frac{\partial S}{\partial r_h}\right|_\Omega = 
\frac{S}{r_h} (D-2 + 2\Sigma_1^-).
\eeq
Combining these we immediately arrive at equation (\ref{eq:COmega}) in the text,
\beq C_\Omega=  -\frac{4\pi r_h T S (D-2+2\,\Sigma_1^-)}
{(D-3 + 2\,\Sigma_1^-)}.
\eeq
This generalises the $D=4$ case derived in \cite{MPS2} to arbitrary $D$.

\section{Isothermal moment of inertia \label{app:I_T}}

To calculate the isothermal moment of inertia tensor 
\beq
\Inertia_T^{ij}= \left( \frac{\partial J^i}{\partial \Omega_j}\right)_T
\eeq
in asymptotically flat Myers-Perry space-times
our starting point is again
\beq T=\frac{D-3-2\Sigma_1^+}{4\pi r_h},\eeq
from which we find
\beq
\left.{d r_h}\right|_T = -\frac{1}{\pi T} 
\left.\frac {\J_i d\J_i }{(1+\J_i^2)^2} \right|_T.
\eeq
Now use this in (\ref{eq:MJdef}), re-written using (\ref{eq:mudef}) in the form 
\beq
J^i=\frac{\varpi r_h^{D-2}}{8\pi}\prod_{k=1}^\dCSA (1+\J_k^2) \J_i ,
\eeq
to deduce that 
\beq
\left. \frac{\partial J^i}{\partial \J_k}\right|_T
=\frac{2 M r_h }{D-2}%\frac{\varpi r_h^{D-2}}{8\pi} \prod_{l=1}^\dCSA (1+\J_l)
\left( \delta_{ik} + 2\left[\frac{(D-3-2\,\Sigma_1^+)(1+\J_k^2) -2(D-2)
}{(D-3-2\,\Sigma_1^+)(1+\J_k^2)^2}\right]\J_i \J_k\right).\label{eq:dJdj}
\eeq
Similar manipulations applied to (\ref{eq:Omegadef}) produce
\beq
\left. \frac{\partial \Omega_j}{\partial \J_k}\right|_T
=\frac{1}{r_h} 
\left\{\frac{(1-\J_j^2)}{(1+\J_j^2)^2} \delta_{jk} + \frac{4\J_j\J_k}
{(D-3-2\Sigma_1^+)(1+\J_j^2)(1+\J_k^2)^2}\right\},\label{eq:dOmegadj}
\eeq
the inverse of which is
\beq
\left. \frac{\partial \J_k}{\partial \Omega_j}\right|_T
=r_h
\left\{\frac{(1+\J_k^2)^2}{(1-\J_k^2)} \delta_{kj} - \frac{4(1+\J_k^2)}
{(\X+4\,\overline\Sigma)(1-\J_j^2)(1-\J_k^2)}\J_k\J_j\right\}.\label{eq:djdOmega}
\eeq
where 
\beq \overline\Sigma:=\sum_{i=1}^\dCSA \frac{\J_i^2}{1-\J_i^4}=\frac 1 2 (\Sigma_1^+ + \Sigma_1^-).\eeq
Equations (\ref{eq:dJdj})--(\ref{eq:djdOmega}) are now easily combined to give
the symmetric isothermal moment of inertia tensor
\beq
\Inertia_T^{ij}= \frac{2 M r_h^2}{D-2}\left\{
\frac{(1+\J_i^2)^2}{(1-\J_i^2)}\delta_{ij} - 
2\J_i \J_j
\left(1+\frac{4}{\vline height 11pt depth 0pt width 0pt \overline \X(1-\J_i^2)(1-\J_j^2)}\right)
\right\},
\eeq
where
\beq
\overline \X:=D-3 + 2\,\Sigma_1^- = t + 4 \overline\Sigma.
\label{eq:tbardef}
\eeq  This is equation (\ref{eq:IT}) in the text.

The determinant of $\Inertia_T$ can be evaluated by observing that the components 
of matrix are of the form
\beq \Inertia_T^{ij}=\frac{2 M r_h^2}{D-2}\bigl( A_i \delta_{ij} - B_i B_j - C_i C_j\bigr)\label{app:ABCmatrix} 
\eeq
with 
\beq A_i=\frac{(1+\J_i^2)^2}{(1-\J_i^2)}, 
\qquad B_i = \sqrt{2}\J_i \qquad \hbox{and} \qquad
C_i= \sqrt{\frac{8}{\overline \X}}\frac {\J_i}{(1-\J_i^2)}.\label{app:ABC}
\eeq
The determinant of (\ref{app:ABCmatrix}) has a compact expression, because the off-diagonal entries factorise,
\bea
\det(\Inertia_T)& =& \\
&&\kern -70pt \left(\frac{2 M r_h^2}{D-2}\right)^\dCSA 
\left(\prod_{i=1}^\dCSA A_i\right)
\left\{1-\sum_{i=1}^\dCSA \frac{\bigl(B_i^2 + C_i^2\bigr)}{A_i}
+ \left(\sum_{i=1}^\dCSA \frac{B_i^2}{A_i}\right) \left(\sum_{j=1}^\dCSA \frac{C_j^2}{A_j}\right)  
 -  \left(\sum_{i=1}^\dCSA \frac{B_i C_i}{A_i}\right)^2\right\}.\nonumber
\eea
A little manipulation, using (\ref{app:ABC}), then yields
\beq
\det\Inertia_T=-\left(\frac{2 M r_h^2}{D-2}\right)^\dCSA 
\frac{\bigl[\X^2-(D-2)(\X-4\Sigma_2^+)\bigr]}{\overline\X}
\prod_{i=1}^\dCSA \frac{(1+\J_i^2)^2}{(1-\J_i^2)}, 
\eeq
with $\overline \X$ defined in (\ref{eq:tbardef}).
This, together with the expression for $C_J$ in (\ref{eq:CJ}), leads to the
compact expression
\beq
\beta C_J \det{\Inertia_T} = -\frac{8\pi^2(D-2)} {(D-3+2\Sigma_1^-)} 
 \left(\frac{2 M r_h^2}{D-2}\right)^{\dCSA+1}
\prod_{i=1}^\dCSA \frac{(1+\J_i^2)^2}{(1-\J_i^2)}.
\label{app:CJdetI}
\eeq

\section{Isentropic moment of inertia \label{app:I_S}}

To calculate the isentropic moment of inertia in asymptotically flat Myers-Perry space-times, again re-write (\ref{eq:Sdef}) as
\beq
S=\frac{\varpi r_h^{D-2}}{4}\prod_{i=1}^\dCSA (1+\J_i^2)
\eeq
from which
\beq
d r_h \bigr|_S = -\frac{2 r_h}{D-2} \sum_{k=1}^\dCSA \left.\frac{\J _k\, d\J_k}{1+\J_k^2}\right|_S,
\eeq
with $dJ_i\bigr|_S = \frac{S}{2\pi}d\J_i\bigr|_S$.
Then
\beq \Omega_i = \frac{\J_i}{r_h(1+\J_i^2)}
\eeq
yields
\beq
\bigl(\Inertia_S^{-1}\bigr)_{ij}=\left(\frac{\partial \Omega_i}{\partial J_j}\right)_S = 
\frac{2\pi}{r_h S}
\left\{
\frac{(1-\J_i^2)}{(1+\J_i^2)^2}\delta_{ij}
+\frac{2\J_i \J_j}{(D-2)(1+\J_i^2)(1+\J_j^2)}
 \right\},\label{app:ISInverse}
\eeq
which was reported in \cite{C-MP}.
Equation (\ref{app:ISInverse}) is easily inverted to give
\beq
\Inertia_S^{ij}= \frac{2 M r_h^2}{D-2}\left\{
\frac{(1+\J_i^2)^2}{(1-\J_i^2)}\delta_{ij} - \frac{2\J_i \J_j}{(D-2+2\Sigma_1^-)}
\frac{(1+\J_i^2)(1+\J_j^2)}{(1-\J_i^2)(1-\J_j^2)}
\right\}.
\eeq
Similar manipulations to those of appendix \ref{app:I_T},
with
\beq A_i=\frac{(1+\J_i^2)^2}{(1-\J_i^2)}, 
\qquad B_i = \sqrt{\frac{2}{\overline \X}} \frac{(1+\J_i^2)}{(1-\J_i^2)}\J_i \qquad \hbox{and} \qquad
C_i= 0,\label{app:tildeABC}
\eeq
reveal that
\beq
\det{\Inertia_S}=
\left(\frac{2 M r_h^2}{D-2}\right)^\dCSA 
\frac{(D-2)}{(D-2 +2\Sigma_1^-)}
\prod_{i=1}^\dCSA \frac{(1+\J_i^2)^2}{(1-\J_i^2)}. \label{app:detIS}
\eeq
Combining this with $C_\Omega$ in (\ref{eq:COmega}) leads to the same 
expression as (\ref{app:CJdetI}),
\beq
\beta C_\Omega \det\Inertia_S = -\frac{8\pi^2(D-2)}{(D-3+2\Sigma_1^-)} \left(\frac{2 M r_h^2}{D-2}\right)^{\dCSA+1}
\prod_{i=1}^\dCSA \frac{(1+\J_i^2)^2}{(1-\J_i^2)},
\label{app:COmegadetI}
\eeq
so (\ref{eq:detITdetIS}) is indeed satisfied.

\end{appendix}

\end{document}